\documentclass[aps,twocolumn,prb,floatfix,showpacs]{revtex4}
\usepackage{amsmath,amssymb}
\usepackage{graphicx}
\usepackage{dcolumn}
\usepackage{bm}
\usepackage{latexsym}
\usepackage{enumerate}

\newcommand{\ASCd}{ASC$_{\dn}$}
\newcommand{\ASCm}{ASC$_-$}
\newcommand{\ASCp}{ASC$_+$}
\newcommand{\ASCpm}{ASC$_{\pm}$}
\newcommand{\ASCu}{ASC$_{\up}$}
\newcommand{\ASCud}{ASC$_{\up,\dn}$}

\newcommand{\band}{\text{band}}

\newcommand{\bH}{\bar{H}}
\newcommand{\boson}{\text{boson}}
\newcommand{\bk}{\mathbf{k}}

\newcommand{\Cc}{C$_\text{C}$}
\newcommand{\Cd}{C$_{\dn}$}
\newcommand{\Cm}{C$_-$}
\newcommand{\Cp}{C$_+$}
\newcommand{\Cpm}{C$_{\pm}$}
\newcommand{\Cs}{C$_\text{S}$}

\newcommand{\Cud}{C$_{\up,\dn}$}
\newcommand{\dd}{\delta_d}
\newcommand{\dn}{\downarrow}

\newcommand{\E}{\mathcal{E}}
\newcommand{\Ed}{\veps_d}
\newcommand{\Edbar}{\bar{\veps}_d}
\newcommand{\eff}{\text{eff}}
\newcommand{\Ek}{\veps_{\bk}}
\newcommand{\Ep}{\veps_p}
\newcommand{\Gam}{\bm{\Gamma}}
\renewcommand{\H}{\hat{H}}
\newcommand{\half}{{\textstyle\frac{1}{2}}}
\newcommand{\imp}{\text{imp}}
\newcommand{\impband}{\text{imp-band}}

\newcommand{\impboson}{\text{imp-boson}}
\newcommand{\loc}{\text{loc}}
\newcommand{\n}{\hat{n}}
\newcommand{\nb}{\bar{n}_b}
\newcommand{\p}{\perp}
\newcommand{\pdag}{\phantom{\dag}}
\newcommand{\s}{\sigma}
\newcommand{\sgn}{\text{sgn}}
\newcommand{\tD}{\tilde{D}}

\newcommand{\tE}{\tilde{E}}
\newcommand{\tEd}{\tilde{\veps}_d}
\newcommand{\tG}{\tilde{\Gamma}}
\newcommand{\tU}{\tilde{U}}

\newcommand{\Ueff}{U_{\eff}}
\newcommand{\Ubar}{\bar{U}}
\newcommand{\up}{\uparrow}
\newcommand{\veps}{\varepsilon}

\begin{document}
\title{Quantum phase transitions in a pseudogap Anderson-Holstein model}

\author{Mengxing Cheng}
\email{mxcheng@phys.ufl.edu}
\affiliation{Department of Physics, University of Florida,
Gainesville, Florida 32611-8440, USA}

\author{Kevin Ingersent}
\affiliation{Department of Physics, University of Florida,
Gainesville, Florida 32611-8440, USA}

\date{\today}

\begin{abstract}
We study a pseudogap Anderson-Holstein model of a magnetic impurity level that
hybridizes with a conduction band whose density of states vanishes in power-law
fashion at the Fermi energy, and couples, via its charge, to a nondispersive
bosonic mode (e.g., an optical phonon). The model, which we treat using
poor-man's scaling and the numerical renormalization group, exhibits quantum
phase transitions of different types depending on the strength of the
impurity-boson coupling. For weak impurity-boson coupling, the suppression of
the density of states near the Fermi energy leads to quantum phase transitions
between strong-coupling (Kondo) and local-moment phases. For sufficiently
strong impurity-boson coupling, however, the bare repulsion between a pair of
electrons in the impurity level becomes an effective attraction, leading to
quantum phase transitions between strong-coupling (charge-Kondo) and
local-charge phases. Even though the Hamiltonian exhibits different symmetries
in the spin and charge sectors, the thermodynamic properties near the two types
of quantum phase transition are closely related under spin-charge interchange.
Moreover, the critical responses to a local magnetic field (for small
impurity-boson coupling) and to an electric potential (for large impurity-boson
coupling) are characterized by the same exponents, whose values place these
quantum critical points in the universality class of the pseudogap Anderson model.
One specific case of the pseudogap Anderson-Holstein model may be realized in a
double-quantum-dot device, where the quantum phase transitions manifest
themselves in the finite-temperature linear electrical conductance.
\end{abstract}

\pacs{71.10.Hf, 75.30.Hx, 72.10.Di, 64.60.ae}
\maketitle

\section{Introduction}
\label{sec:intro}

Quantum phase transitions (QPTs) take place between competing ground states at
the absolute zero of temperature ($T=0$) upon variation of a nonthermal control
parameter.\cite{Sondhi:97,Sachdev:99} QPTs are thought to play a role in many
important open problems in condensed-matter physics, including
high-temperature superconductivity,\cite{Anderson:87,Dagotto:94,Broun:08}
the phase diagram for magnetic heavy-fermion metals,\cite{Gegenwart:08,Si:10}
and various types of metal-insulator
transition.\cite{Giamarchi:08,Dobrosavljevic:11}

An interesting class of zero-temperature transitions is impurity or boundary
QPTs at which only a subset of system degrees of freedom becomes
critical.\cite{Vojta:06} A well-studied example arises in the pseudogap
Anderson impurity model\cite{Withoff:90,Gonzalez-Buxton:96,Bulla:97,%
Gonzalez-Buxton:98,Logan:00,Bulla:00,Glossop:03} of
an interacting impurity level hybridizing with a host density of states that
vanishes in power-law fashion precisely at the Fermi energy---a property
that can be realized in a number of systems including unconventional
\emph{d}-wave superconductors,\cite{Sigrist:91} certain semiconductor
heterostructures,\cite{Volkov:85} and a particular double-quantum-dot
setup.\cite{Dias:06,Dias:08,Dias:09} The reduction of the density of states
near the Fermi energy leads to QPTs between strong-coupling (Kondo-screened)
and local-moment phases.\cite{Withoff:90} At the transitions, the system
exhibits a critical response to a local magnetic field applied only to the
impurity site.\cite{Ingersent:02,Kircan:04,Fritz:04}

Impurity quantum phase transitions have been predicted\cite{Hofstetter:01,%
Oreg:03,Pustilnik:04,Galpin:05,Zitko:06,Dias:06,Dias:08,Dias:09,Logan:09,%
Wong:12} and possibly observed\cite{Potok:07,Roch:08} to arise in various
nanodevices.
While strong electron-electron interactions are an integral element of such
nanodevices, experiments on single-molecule transistors\cite{Park:00} and
quantum-dot cavities\cite{Weig:04} have also highlighted the importance of
electron-phonon interactions. The main aspects of the last two
experiments appear to be captured by variants of the Anderson-Holstein
model, which supplements the Anderson model\cite{Anderson:61} for a
magnetic impurity in a metallic host with a Holstein
coupling\cite{Holstein:59} of the impurity charge to a local bosonic mode,
usually assumed to represent an optical phonon.
The model has been studied since the 1970s in connection with the mixed-valence
problem,\cite{Kaga:80,Schonhammer:84,Alascio:88,Ostreich:91,Hewson:02,Jeon:03,%
Zhu:03,Lee:04} the role of negative-$U$ centers in
superconductors,\cite{Simanek:79,Ting:80} and most recently, single-molecule
devices.\cite{Cornaglia:04,Cornaglia:05} Various analytical approximations as
well as nonperturbative numerical renormalization-group calculations have shown
that the Holstein coupling reduces the Coulomb repulsion between two electrons
in the impurity level, even yielding effective electron-electron attraction
for sufficiently strong impurity-boson coupling. Furthermore, for the full
Anderson-Holstein model with nonzero hybridization, as the impurity-boson
coupling increases from zero, there can be a crossover from a conventional
Kondo effect, involving conduction-band screening of the impurity spin degree
of freedom, to a ``charge Kondo effect'' in which it is the impurity ``isospin''
or deviation from half-filling that is quenched by the conduction band.
However, the evolution between these limits is entirely smooth, and the model
exhibits no QPT.

This paper reports the results of study of a pseudogap Anderson-Holstein model
which incorporates the structured conduction-band density of states from the
pseudogap Anderson model into the Anderson-Holstein model. The essential physics
of the problem, revealed using a combination of poor-man's scaling and the
numerical renormalization group (NRG), is shown to depend on the sign of the effective
Coulomb interaction between two electrons in the impurity level, on the presence
or absence of particle-hole (charge-conjugation) symmetry and time-reversal
symmetry, and on the value of the exponent $r$ characterizing the variation
$\rho(\veps)\propto|\veps|^r$ of the density of states near the Fermi energy
$\veps=0$. Even though the pseudogap Anderson-Holstein Hamiltonian has a lower
symmetry than the pseudogap Anderson Hamiltonian, the universal properties of
the former model, including the structure of the renormalization-group fixed points
and the values of critical exponents describing properties in the vicinity of
those fixed points, are identical to those of the latter model as generalized
to allow for negative (attractive) as well as positive values of the local
interaction $U$ between two electrons in the impurity level.\cite{Fritz:04} The
pseudogap Anderson-Holstein model can therefore be regarded in part as
providing a physically plausible route to accessing the negative-$U$ regime of
the pseudogap Anderson model. Anderson impurities with $U<0$ have recently
attracted attention as a possible route to achieving enhanced thermoelectric
power.\cite{Andergassen:11}

The remainder of this paper is organized as follows: Section \ref{sec:model}
introduces the pseudogap Anderson-Holstein model, analyzes special cases in
which the model reduces to problems that have been studied previously, outlines
a perturbative scaling analysis of the full model, and summarizes the numerical
renormalization-group approach used to provide nonperturbative solutions of the
model. Section \ref{sec:small-r_symm} presents results under conditions of
particle-hole and time-reversal symmetry while Sec.\ \ref{sec:small-r}
addresses the general model with band exponent $r$ between $0$ and $1$. Section
\ref{sec:r=2} focuses on the specific case $r=2$ relevant to a boson-coupled
double-quantum-dot device. Section \ref{sec:summary} summarizes the main results
of the paper. The Appendix \ref{sec:scaling} contains details of the perturbative
scaling analysis.

\section{Model Hamiltonian, Preliminary Analysis, and Solution Method}
\label{sec:model}

\subsection{Pseudogap Anderson-Holstein model}

In this work, we study the pseudogap Anderson-Holstein model described by the
Hamiltonian
\begin{equation}
\label{H_PAHM}
\H = \H_{\imp} + \H_{\band} + \H_{\boson} + \H_{\impband} + \H_{\impboson} ,
\end{equation}
where
\begin{subequations}
\begin{align}
\label{H_imp}
\H_{\imp}&
= \dd \, (\n_d-1) + \half U(\n_d-1)^2 , \\
\label{H_band}
\H_{\band}&
= \sum_{\bk,\s}\veps^{\pdag}_{\bk}c^{\dag}_{\bk\s}
  c^{\pdag}_{\bk \s} , \\
\label{H_boson}
\H_{\boson}&
= \omega_0 \, b^{\dag} b, \\
\label{H_impband}
\H_{\impband}&
=  \frac{1}{\sqrt{N_k}} \sum_{\bk,\s}
   \bigl( V_{\bk} d^{\dag}_{\s}c^{\pdag}_{\bk\s}+ \text{H.c.}), \\
\label{H_impboson}
\H_{\impboson}&
= \lambda(\n_d-1)(b + b^{\dag}) .
\end{align}
\end{subequations}
Here, $d_{\s}$ annihilates an electron of spin $z$ component
$\s=\pm\half$ (or $\s=\:\up,\,\dn$) and energy
$\Ed=\dd-\half U<0$ in the impurity level, $\n_d = \sum_{\s} \n_{d\s}$
(with $\n_{d\s} = d_{\s}^{\dag} d_{\s}$) is the total impurity
occupancy, and $U>0$ is the Coulomb repulsion between two electrons
in the impurity level.\cite{H_imp:alt} $V_{\bk}$ is the hybridization matrix
element between the impurity and a conduction-band state of energy $\Ek$
annihilated by fermionic operator $c_{\bk\s}$, and $\lambda$ characterizes
the Holstein coupling of the impurity occupancy to the displacement of a local
vibrational mode of frequency $\omega_0$. $N_k$ is the number of unit cells in
the host metal and, hence, the number of inequivalent $\bk$ values. Without loss
of generality, we take $V_{\bk}$ and $\lambda$ to be real and non-negative. For
compactness of notation, we drop all factors of the reduced Planck constant
$\hbar$, Boltzmann's constant $k_B$, the impurity magnetic moment $g\mu_B$,
and the electronic charge $e$.

The conduction-band dispersion $\Ek$ and the hybridization $V_{\bk}$
affect the impurity degrees of freedom only through the hybridization
function\cite{Gamma-notation}
\begin{equation}
\label{Gamma:def}
\Gam(\veps) \equiv \frac{\pi}{N_k} \sum_{\bk}
   |V_{\bk}|^2 \delta(\veps-\Ek) .
\end{equation}
To focus on the most interesting physics of the model, we assume a simplified
form
\begin{equation}
\label{Gamma:power}
\Gam(\veps) = \Gamma \, |\veps/D|^r \, \Theta(D-|\veps|) ,
\end{equation}
where $\Theta(x)$ is the Heaviside function and we refer to the prefactor
$\Gamma$ as the hybridization width. In this notation, the case $r=0$
represents a conventional metallic hybridization function. This paper focuses
on cases $r>0$ in which the hybridization function exhibits a power-law pseudogap
around the Fermi energy. One way that such a hybridization function can arise is
from a purely local hybridization matrix element $V_{\bk}=V$ combined with a
density of states (per unit cell per spin orientation) varying as
\begin{equation}
\label{rho:def}
\rho(\veps) \equiv N_k^{-1} \sum_{\bk}\delta(\veps-\Ek)
= \rho_0|\veps/D|^r\Theta(D-|\veps|),
\end{equation}
in which case $\Gamma=\pi\rho_0 V^2$.
However, the results presented in this paper apply equally to situations in which the
$\bk$ dependence of the hybridization contributes to the energy dependence of
$\Gam(\veps)$.

The assumption that $\Gam(\veps)$ exhibits a pure power-law dependence
over the entire width of the conduction band is a convenient idealization. More
realistic hybridization functions in which the power-law variation is restricted to a
region around the Fermi energy exhibit the same qualitative physics, with
modification only of nonuniversal properties such as critical couplings and
Kondo temperatures.

The properties of the Hamiltonian specified by Eqs.\
\eqref{H_PAHM}--\eqref{Gamma:power} turn out to depend crucially on whether or
not the system is invariant under the particle-hole transformation
$c_{\bk\s}^{\pdag}\to c_{\bk\s}^{\dag}$,
$d_{\s}^{\pdag}\to -d_{\s}^{\dag}$, $b\to -b$, which maps
$\Ek\to -\Ek$ and $\dd\to -\dd$. For the symmetric
hybridization function given in Eq.\ \eqref{Gamma:power}, the condition for
particle-hole symmetry is $\dd=0$ corresponding to $\Ed=-\half U$.

\subsection{Review of related models}
\label{subsec:special-cases}

Before addressing the full pseudogap Anderson-Holstein model, it is useful to
review two limiting cases that have been studied previously.

\subsubsection{Pseudogap Anderson model}
\label{subsubsec:PAM}

For coupling $\lambda=0$, the pseudogap Anderson-Holstein model reduces to the
pseudogap Anderson model\cite{Gonzalez-Buxton:96,Bulla:97,Gonzalez-Buxton:98,%
Logan:00,Bulla:00,Glossop:03,Kircan:04,Fritz:04} plus free local bosons. In the
conventional ($r=0$) Anderson impurity model, the generic low-temperature limit
is a strong-coupling regime in which the impurity level is effectively absorbed
into the conduction band.\cite{Krishna-murthy:80} In the pseudogapped ($r>0$)
variant of the model, the depression of the hybridization function around the Fermi
energy gives rise to a competing local-moment phase in which the impurity
retains an unscreened spin degree of freedom all the way to absolute zero.
The $T=0$ phase diagram of this model depends on the presence or absence of
particle-hole symmetry\cite{Gonzalez-Buxton:98} and of time-reversal
symmetry.\cite{Fritz:04}

\noindent
\textit{Behavior at particle-hole symmetry} ($\dd=0$):
For any band exponent $0<r<\half$, in zero magnetic field there is a continuous
QPT at a critical coupling $\Gamma=\Gamma_c(r,U,\dd=0)$ between the
local-moment phase and a symmetric strong-coupling phase. In the local-moment
phase (reached for $0\le\Gamma<\Gamma_c$), the impurity
contributions\cite{impurity-props} to the entropy and to the static spin
susceptibility approach the low-temperature limits $S_{\imp}=\ln 2$ and
$T\chi_{s,\imp}=1/4$, respectively, while conduction electrons at the Fermi
energy experience an $s$-wave phase shift $\delta_0(\veps=0) = 0$.
In the symmetric strong-coupling phase ($\Gamma>\Gamma_c$), the corresponding
properties are $S_{\imp}=2r\ln 2$, $T\chi_{s,\imp}=r/8$, and $\delta_0(0) =
-(1-r)(\pi/2)\,\sgn\,\veps$, all indicative of partial quenching of the
impurity degrees of freedom.

A magnetic field that couples to the band electrons moves the zero in the
density of states of each spin species away from the Fermi level and washes
out all pseudogap physics at energies below the Zeeman scale.\cite{Fritz:04}
More interesting is the breaking of time-reversal symmetry by a local
magnetic field that couples only to the impurity degree of freedom and
enters the Anderson model through an additional Hamiltonian term
\begin{equation}
\label{h:def}
H_h = \frac{h}{2} \, (\n_{d\up}-\n_{d\dn}).
\end{equation}
The critical response to an infinitesimal $h$ reveals that the transition
between the local-moment and strong-coupling phases takes place at an
interacting quantum critical point.\cite{Ingersent:02,Kircan:04,Fritz:04}
However, a finite value of $h$ destabilizes both the local-moment
phase\cite{Fritz:04} and the symmetric strong-coupling
phase,\cite{SSC-mag-field} and destroys the QPT between the
two.\cite{Fritz-note} For any $h\ne 0$, the ground state of the
particle-hole-symmetric model (with $U>0$) is a fully-polarized
local moment that is asymptotically decoupled from the conduction
band.\cite{Fritz-note}

For $r\ge\half$, the symmetric strong-coupling fixed point is unstable
even in zero magnetic field,\cite{Bulla:97,Gonzalez-Buxton:98} and a
particle-hole-symmetric system lies in the local-moment phase for all values
of $\Gamma$.

\begin{figure}[t]
\centering
\includegraphics[width=.95\columnwidth]{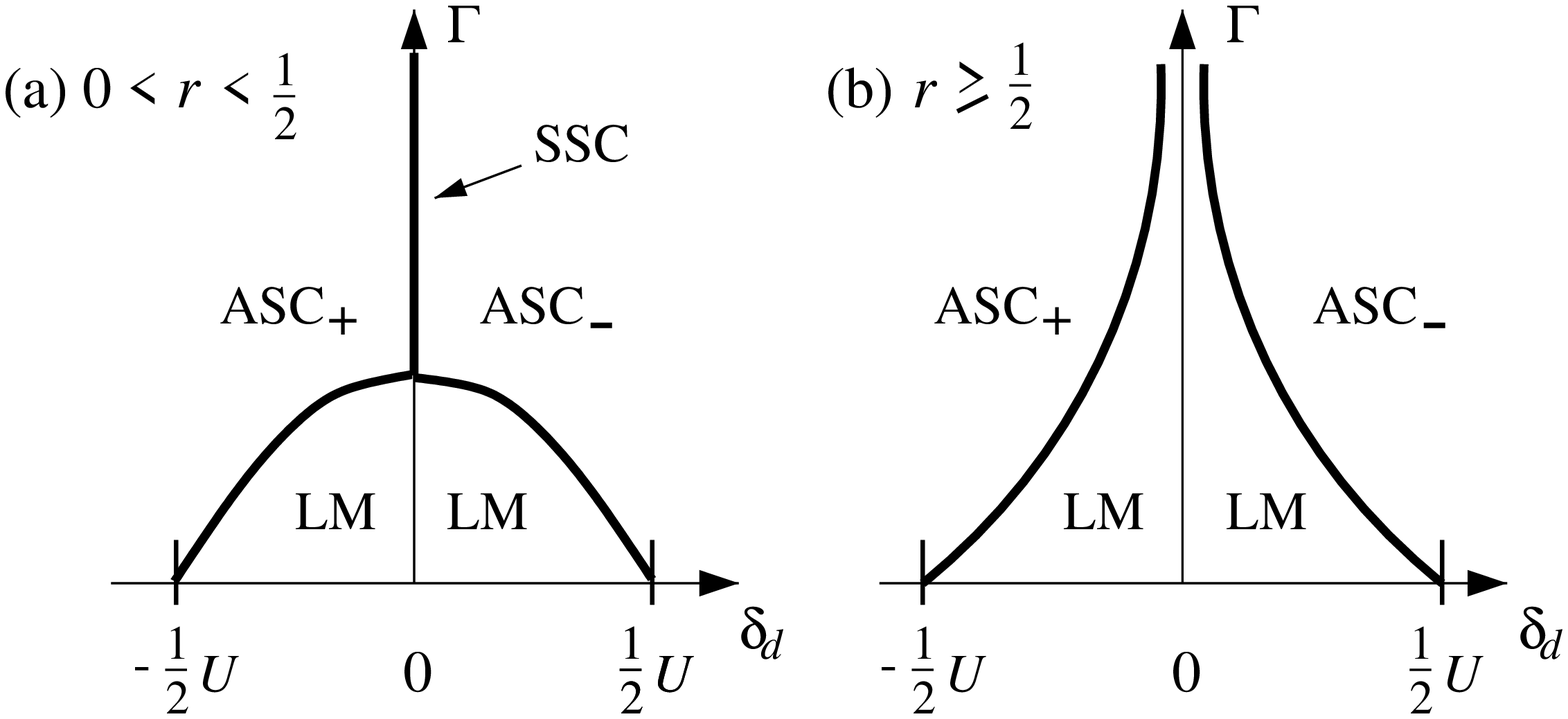}
\caption{\label{fig:PAM_phase}
Schematic $\dd$-$\Gamma$ phase diagrams of the pseudogap Anderson model
[Eqs.\ \protect\eqref{H_PAHM}--\protect\eqref{Gamma:power} with $\lambda=0$]
for band exponents (a) $0 < r < \half$, (b) $r\ge\half$. Generically, the
system falls into either a local-moment phase (LM) or one of two asymmetric
strong-coupling phases (\ASCpm). However, there is also a symmetric
strong-coupling phase (the line labeled SSC) that is reached only for
$0<r<\half$ under conditions of strict particle-hole symmetry ($\dd=0$)
and for sufficiently large hybridization widths $\Gamma$.}
\end{figure}

\noindent
\textit{Behavior away from particle-hole symmetry} ($\dd \ne 0$):
In zero magnetic field, the model remains in the local-moment phase described
above for all $|\dd|<\half U$ (i.e., $-U < \Ed < 0$) and
$\Gamma < \Gamma_c(r,U,\dd)\equiv \Gamma_c(r,U,-\dd)$.
As shown schematically in Fig.\ \ref{fig:PAM_phase}, the critical hybridization
width $\Gamma_c$ increases monotonically from zero as $|\dd|$ drops below
$\half U$. For $0<r<\half$ [Fig.\ \ref{fig:PAM_phase}(a)],
$\Gamma_c(r,U,\dd)$ smoothly approaches the symmetric critical value
$\Gamma_c(r,U,0)$ as $\dd\to 0$. For $r\ge\half$ [Fig.\
\ref{fig:PAM_phase}(b)], $\Gamma_c(r,U,\dd)$ instead diverges as
$\dd\to 0$, consistent with the $\dd=0$ behavior discussed above.

For $\dd\ne 0$ and $\Gamma > \Gamma_c$, the model lies in one of two
asymmetric strong-coupling phases that share the low-temperature properties
$S_{\imp}=0$ and $T\chi_{s,\imp}=0$. For $\dd>0$, the Fermi-energy
phase shift is $\delta_0(0) = -\pi\,\sgn\,\veps$, while the ground-state
charge (total fermion number measured from half-filling) is $Q=-1$. For
$\dd<0$, by contrast, $\delta_0(0) = +\pi\,\sgn\,\veps$ and $Q=+1$. We
label these two phases \ASCm\ and \ASCp\ according to the sign of $Q$.

For $r<r^*\simeq 3/8$, the low-temperature physics on the phase boundary
$\Gamma=\Gamma_c(r,U,\dd\ne 0)$ is identical to that at
$\Gamma=\Gamma_c(r,U,0)$, whereas for $r>r^*$ the properties are
distinct.\cite{Gonzalez-Buxton:98,Fritz:04} For $r^* < r < 1$, the response to
an infinitesimal local magnetic field shows that asymmetric transitions take
place at two interacting quantum critical points (one for $\dd>0$, the other
for $\dd<0$). For $r\ge 1$, the QPTs are first-order\cite{Ingersent:02} and
can be interpreted as renormalized level crossings between the local-moment
doublet and the \ASCpm\ singlet ground states.\cite{Fritz:04}

The asymmetric strong-coupling phase is stable over a range of local magnetic
fields.\cite{SSC-mag-field} However, at a critical value of $|h|$ the system
undergoes a level-crossing QPT into the same fully polarized phase as is found
at particle-hole symmetry.\cite{Fritz:04}

\noindent
\textit{Relationship to the pseudogap Kondo model}:
In cases where $\Gamma\ll\half U-|\dd|$, the pseudogap Anderson model can
be mapped via a Schrieffer-Wolff transformation\cite{Schrieffer:66} onto the
pseudogap Kondo model.\cite{Withoff:90} The latter model exhibits QPTs entirely
equivalent to those described above.\cite{Gonzalez-Buxton:98}  This allows us
to identify critical exponents obtained previously for the pseudogap Kondo
model\cite{Ingersent:02} as the values that apply to the special case
$\lambda=0$ of the pseudogap Anderson-Holstein model.

\subsubsection{Anderson-Holstein model}
\label{subsubsec:AHM}

For $r=0$, the pseudogap Anderson-Holstein model reduces to the
Anderson-Holstein model.\cite{Simanek:79,Ting:80,Kaga:80,Schonhammer:84,%
Alascio:88,Schuttler:88,Ostreich:91,Hewson:02,Jeon:03,Zhu:03,Lee:04,%
Cornaglia:04,Cornaglia:05} Insight into the physics of both models can be
gained by performing a canonical transformation of the Lang-Firsov
type\cite{Lang:62} to eliminate the Holstein coupling between the bosons
and the impurity occupancy [Eq.\ \eqref{H_impboson}].
The transformation\cite{Hewson:02}
\begin{equation}
\label{Lang-Firsov}
\bH=e^S \H e^{-S} \quad \text{with} \quad
S=\frac{\lambda}{\omega_0} (\n_d - 1) \bigl(b^{\dag}-b\bigr)
\end{equation}
maps Eq.\ \eqref{H_PAHM} to
\begin{equation}
\label{bar H_PAHM}
\bH=\bH_{\imp}+\H_{\band}+\H_{\boson}+\bH_{\impband},
\end{equation}
in which $\H_{\boson}$ and $\H_{\band}$ remain as given in Eqs.\ \eqref{H_boson}
and \eqref{H_band}, respectively. $\bH_{\imp}$ is identical to
$\H_{\imp}$ [Eq.\ \eqref{H_imp}] apart from the replacement of $U$ by
\begin{equation}
\label{Ubar:def}
\Ubar=U-2\,\Ep,
\end{equation}
where the polaron energy
\begin{equation}
\label{Ep:def}
\Ep=\lambda^2/\omega_0
\end{equation}
represents an important energy scale in the problem.
The invariance of $\dd$ under the mapping implies a renormalization of the
level energy from $\Ed=\dd-\half U$ to
\begin{equation}
\label{Edbar:def}
\Edbar = \Ed + \Ep.
\end{equation}
Finally, the impurity-band coupling term becomes
\begin{equation}
\label{bar H_impband}
\bH_{\impband}=\frac{1}{\sqrt{N_s}}
 \sum_{\bk,\s}V_{\bk}^{\pdag}\Bigl(B^{\dag}d_{\s}^{\dag} c_{\bk\s}^{\pdag}
 +\text{H.c} \Bigr),
\end{equation}
with
\begin{equation}
\label{B:def}
B=e^{-(\lambda/\omega_0) (b^{\dag}-b)} .
\end{equation}

The canonical transformation in Eq.\ \eqref{Lang-Firsov} maps the local boson
mode to
\begin{equation}
\label{bar b:def}
\bar{b} = e^S b \, e^{-S} = b - (\lambda/\omega_0) (\n_d - 1),
\end{equation}
effectively defining a different displaced-oscillator basis for each value of
the impurity occupancy $n_d$, namely, the basis that minimizes the ground-state
energy of $\H_{\imp}+\H_{\boson}+\H_{\impboson}$. The elimination of the Holstein
coupling is accompanied by two compensating changes to the Hamiltonian: a
reduction in the magnitude---or even a change in the sign---of the interaction
within the impurity level, reflecting the fact that Eq.\ \eqref{H_impboson}
lowers the energy of the empty and doubly occupied impurity configurations
relative to single occupation; and incorporation into the impurity-band term
[Eq.\ \eqref{bar H_impband}] of operators $B$ and $B^{\dag}$ that cause each
hybridization event to be accompanied by the creation and absorption of a
packet of bosons as the local mode adjusts to the change in the impurity
occupancy $n_d$.

The analysis of Eq.\ \eqref{bar H_PAHM} is trivial in the case $\Gamma=0$ of
zero hybridization where the Fock space can be partitioned into subspaces of
fixed impurity occupancy $n_d=0$, 1, and 2, and the ground state within each
sector corresponds to the vacuum of the transformed boson mode. It can be seen
from Eq.\ \eqref{Ubar:def} that the effective on-site Coulomb interaction
changes sign at $\lambda=\lambda_0$, where
\begin{equation}
\label{lambda_0:def}
\lambda_0 = \sqrt{\omega_0 U/2} \, .
\end{equation}
For weak bosonic couplings $\lambda<\lambda_0$, the effective interaction is
repulsive, and for $|\dd|<\half \Ubar$ the impurity ground state is a spin
doublet with $n_d = 1$ and $\s = \pm\half$. For $\lambda>\lambda_0$, by
contrast, the strong coupling to the bosonic mode yields an attractive
effective on-site interaction and for $|\dd|<-\half \Ubar$ the two lowest-energy
impurity states are spinless but have a charge (relative to half filling)
$Q = n_d-1 = \pm 1$; these states are degenerate only under conditions of
strict particle-hole symmetry ($\dd = 0$).

Various limiting behaviors of the full Anderson-Holstein model with
$\Gamma\ne 0$ are understood:\cite{Schuttler:88,Hewson:02}
\begin{enumerate}[(i)]
\item
If $\omega_0$ and $\lambda$ are both taken to infinity in such a way that $\Ep$
defined in Eq.\ \eqref{Ep:def} approaches a finite value, the model behaves just
like the pure-fermionic Anderson model with $U$ replaced by $\Ubar$ while
$\Gamma$ and $\dd$ are unaffected by the bosonic coupling (implying that
$\Ed=\dd-\half U$ is replaced by $\Edbar$).
\item
In the \textit{instantaneous} or \textit{anti-adiabatic} limit
$\Gamma\ll\omega_0<\infty$, the bosons adjust rapidly to any change in the
impurity occupancy; for $\omega_0,U\gg\Ep$, the physics
essentially remains that of the Anderson model with $U\to \Ubar$, while for
$\omega_0,U\ll\Ep$, there is also a reduction from $\Gamma$ to
$\Gamma_{\eff}=\Gamma\exp(-\Ep/\omega_0)$ in the hybridization width
describing scattering between the $n_d=0$ and $n_d=2$ sectors, reflecting the
reduced overlap between the ground states in these two sectors.
\item
In the \textit{adiabatic} limit $\Gamma\gg\omega_0$, by contrast, the bosons are
unable to adjust on the typical time scale of hybridization events, and neither
$U$ nor $\Gamma$ undergoes significant renormalization.
\item
In the physically most relevant regime $\Gamma\lesssim\omega_0<U,D$, NRG
calculations\cite{Hewson:02, Cornaglia:04} show that for $\Gamma\ll|\dd|\ll U$,
there is a smooth crossover from a conventional charge-sector Kondo effect for
$\lambda\ll\lambda_0$ (and thus $\Gamma\ll\Ubar$) to a charge-sector analog of
the Kondo effect for $\lambda\gg\lambda_0$ (and $\Gamma\ll-\Ubar$). The primary
goal of the present work is to explain how this physics is modified by the
presence of a pseudogap in the impurity hybridization function.
\end{enumerate}

\subsection{Poor-Man's Scaling}
\label{subsec:scaling}

As a preliminary step in the analysis of the pseudogap Anderson-Holstein
Hamiltonian, we develop poor-man's scaling equations describing the evolution
of model parameters under progressive reduction of the conduction bandwidth.
Haldane's scaling analysis \cite{Haldane:78} of the metallic ($r=0$) Anderson
model in the limit $U\gg D$ has previously been extended to the pseudogap case
$r>0$, both for infinite\cite{Gonzalez-Buxton:96} and finite\cite{Cheng:inprep}
on-site interactions $U$. Here, the analysis is further generalized to treat the
anti-adiabatic regime of the Anderson-Holstein model with both metallic and
pseudogapped densities of states.

Our analysis begins with the Lang-Firsov canonical transformation \eqref{Lang-Firsov}.
In the anti-adiabatic regime, it is a good approximation to calculate all
physical properties in the vacuum state of the transformed boson mode defined
in Eq. \eqref{bar b:def}. We therefore focus on many-body states representing
the direct product of the bosonic vacuum with the half-filled Fermi sea and
with one of the four possible configurations of the impurity level. In the
atomic limit $\Gamma=0$, the energies of the states having impurity occupancy
$n_d=0$, $1$, and $2$ can be denoted $E_0$, $E_1=E_0+\Edbar$, and
$E_2=E_1+\Edbar+\Ubar=2E_1-E_0+\Ubar$, respectively.

The poor-man's scaling procedure involves progressive reduction of the
conduction-band halfwidth from $D$ to $\tD$. At each infinitesimal step
$\tD\to\tD+d\tD<\tD$, the energies $E_0$, $E_1$, and $E_2$, as well as the
hybridization function $\Gamma$ are adjusted to compensate for the elimination of
virtual hybridization processes involving band states in the energy windows
$\tD+d\tD<\veps<\tD$ and $-\tD<\veps<-(\tD+d\tD)$. An added complication in
the Anderson-Holstein model is the presence of the operators $B$ and
$B^{\dag}$ in Eq.\ \eqref{bar H_impband}, which allow virtual excitation of
states having arbitrarily high boson occupation numbers $\nb=\bar{b}^{\dag}
\bar{b}^{\pdag}$. As detailed in the Appendix \ref{sec:scaling},
summation over all such intermediate states leads to the scaling equations
\begin{align}
\label{U:scaling}
\frac{d\tU}{d\tD}
&= \frac{2\tG}{\pi} \biggl[ \frac{1}{\E(\tD\!+\!\tEd)}
   - \frac{1}{\E(\tD\!-\!\tEd)} \notag \\
&\qquad + \frac{1}{\E(\tD\!-\!\tU\!-\!\tEd)}
   - \frac{1}{\E(\tD\!+\!\tU\!+\!\tEd)} \, \biggr] , \\
\label{Ed:scaling}
\frac{d\tEd}{d\tD}
&= \frac{\tG}{\pi} \biggl[ \frac{1}{\E(\tD\!-\!\tEd)}
   - \frac{2}{\E(\tD\!+\!\tEd)}
   + \frac{1}{\E(\tD\!+\!\tU\!+\!\tEd)} \, \biggr] , \\
\label{Gamma:scaling}
\frac{d\tG}{d\tD}
&= r\,\frac{\tG}{\tD} \, ,
\end{align}
for renormalized model parameters that take bare values
$\tU=\Ubar$ [Eq.\ \eqref{Ubar:def}],
$\tEd=\Edbar$ [Eq.\ \eqref{Edbar:def}], and $\tG=\Gamma$ for $\tD=D$.
In these equations, the energy scale
\begin{equation}
\label{E:def}
\E(E) = E \biggl/ S\biggl(\frac{E}{\omega_0}, \,
  \frac{\Ep}{\omega_0}\biggr),
\end{equation}
is defined in terms of a dimensionless function
\begin{equation}
\label{S:def}
S(a, x) = a \, e^{-|x|} \sum_{n = 0}^{\infty} \frac{1}{n!} \, \frac{x^n}{a+n}
\equiv a \, \Gamma(a) \, \gamma^*(a, -x) \, e^{-|x|},
\end{equation}
where $\Gamma(a)$ is the gamma function and $\gamma^*(a,x)$ is related to
the lower incomplete gamma function.\cite{Gradshteyn}

In the case $\lambda=0$ where $\E(E)=E$, Eqs.\
\eqref{U:scaling}--\eqref{Gamma:scaling} reduce to the scaling equations for
the pseudogap Anderson model.\cite{Cheng:inprep}  The pseudogap in the
hybridization function produces a strong downward rescaling of $\tG$
[see Eq.\ \eqref{Gamma:scaling}] that leads, via Eqs.\ \eqref{U:scaling} and
\eqref{Ed:scaling} to weaker renormalization of $\tU$ and $\tEd$ than would
occur in a metallic ($r=0$) host. For $\lambda>0$, one finds that $|\E(E)|>|E|$,
so the bosonic coupling acts to further reduce (in magnitude) the right-hand
sides of Eqs.\ \eqref{U:scaling} and \eqref{Ed:scaling}, and produces still
slower renormalization of $\tU$ and $\tEd$ with decreasing $\tD$.
It should be noted that neither the bosonic energy $\omega_0$ nor the
electron-boson coupling $\lambda$ is renormalized under the scaling procedure,
and that the scaling equations respect particle-hole symmetry in that bare
couplings satisfying $\Ed=-\half U$ inevitably lead to rescaled couplings that
satisfy $\tEd=-\half\tU$.

Equation \eqref{Gamma:scaling} can readily be solved to give
\begin{equation}
\label{tildeGamma}
\tG(\tD) = (\tD/D)^r \; \Gamma.
\end{equation}
For $\lambda>0$, it is not possible to integrate Eqs.\ \eqref{U:scaling} and
\eqref{Ed:scaling} in closed form due to the presence of the nontrivial function
$\E(E)$ on their right-hand sides. The equations have been derived only to
lowest order in nondegenerate perturbation theory, and are therefore limited in
validity to the range$|\tU|,\,|\tEd|,\,\tG\ll\tD$.
Nonetheless, one may be able to obtain useful insight into the qualitative
physics of the model through numerical integration of Eqs.\ \eqref{U:scaling}
and \eqref{Ed:scaling} until one of the following conditions is met, implying
entry into a low-energy regime governed by a simpler effective model than the
full pseudogap Anderson-Holstein model:
\begin{enumerate}[(i)]
\item
If $\tEd,\,\tU+2\tEd>\tD>\tG$, the system should enter the empty-impurity
region of the strong-coupling phase, where the impurity degree of freedom is
frozen with an occupancy close to zero.
\item
If $-(\tU+\tEd),\,-(\tU+2\tEd)>\tD>\tG$, the system should enter the
full-impurity region of the strong-coupling phase, where the impurity degree
of freedom is frozen with an occupancy close to two.
\item
If $-\tEd,\,\tU+\tEd>\tD>\tG$, the system is expected to enter an
intermediate-energy local-moment regime in which the impurity states with
$n_d\ne 1$ are frozen out. As discussed further in Sec.\
\ref{subsec:small-r_symm:boundaries},
one can perform a generalization of the Schrieffer-Wolff
transformation\cite{Schrieffer:66} to map the pseudogap Anderson-Holstein model
to a pseudogap Kondo model with the density of states in Eq.\ \eqref{rho:def}.
Depending on the value of the Kondo exchange coupling generated by the
Schrieffer-Wolff transformation, the system may lie either in the strong-coupling
phase of the pseudogap Kondo model (which should correspond to another region of
the strong-coupling phase of the Anderson-Holstein model) or in a local-moment
phase where the impurity retains a free two-fold spin degree of freedom down to
absolute zero.
\item
If $\tEd,-(\tU+\tEd)>\tD>\tG,\,\tU+2\tEd$, the system should enter an
intermediate-energy local-charge regime in which the $n_d=1$ impurity states
become frozen out. A generalized Schrieffer-Wolff transformation can map the
pseudogap Anderson-Holstein model to a pseudogap charge-Kondo model (see
Sec.\ \ref{subsec:small-r_symm:boundaries}). The system may lie either in the
strong-coupling phase of the charge-Kondo model (yet another region of the
strong-coupling phase of the Anderson-Holstein model) or in a local-charge
phase of both models (where the impurity retains a free two-fold charge degree
of freedom down to absolute zero).
\item
If $\tG>\tD>|\tEd|$ and/or $\tG>\tD>|\tU+\tEd|$, then the system should
enter the mixed-valence region of the strong-coupling phase.
\end{enumerate}

Since each of the crossovers described above lies beyond the range of validity
of the scaling equations, the preceding analysis is only suggestive. In order to
provide a definitive account of the pseudogap Anderson-Holstein model, it is
necessary to obtain full, nonperturbative solutions, such as those provided by
the numerical renormalization group. However, we shall return to the scaling
equations in Sec.\ \ref{subsec:r=2:phase_diagram} to assist in the quantitative
analysis of numerical results.

\subsection{Numerical solution method}

We have solved the model Eq.\ \eqref{H_PAHM} using the numerical
renormalization-group (NRG) method,\cite{Krishna-murthy:80,Wilson:75,Bulla:08} as
extended to treat problems with an energy-dependent hybridization
function,\cite{Bulla:97,Gonzalez-Buxton:98} and ones that involve local
bosons.\cite{Hewson:02} Briefly, the procedure involves three key steps:
(i) Division of the full range of conduction-band energies
$-D \leq \Ek \leq D$ into a set of logarithmic intervals bounded by
$\veps_m=\pm D\Lambda^{-m}$ for $m=0,1,2,\ldots$, where $\Lambda>1$ is the
Wilson discretization parameter. The continuum of states within each
interval is replaced by a single state of each spin $\s$, namely, the
linear combination of states lying within the interval that couples to the
impurity. (ii) Application of the Lanczos procedure to map the discretized
version of $\H_{\band}$ onto a tight-binding form\cite{e_n}
\begin{equation}
\label{tight-binding}
\H_{\band} = \sum_{n=0}^{\infty} \Lambda^{-n/2} \, t_n \sum_{\s}
\bigl( f_{n\s}^{\dag} f_{n-1,\s}^{\pdag} + \text{H.c.}),
\end{equation}
where $f_{0\s}\propto \sum_{\bk} V_{\bk} c_{\bk\s}$, and
$\{ f_{n,\s}^{\pdag}, f_{n',\s'}^{\dag} \} = \delta_{n,n'}\,
\delta_{\s,\s'}$. The hopping parameters $t_n$ (with $t_0=0$) contain all
information about the energy dependence of the hybridization function
$\Gam(\veps)$.
(iii) Iterative solution of the problem via diagonalization of a sequence of
rescaled Hamiltonians
\begin{equation}
\H_0 = \H_{\imp}+\H_{\boson}+\H_{\impband}+\H_{\impboson}-E_{G,0}
\end{equation}
and
\begin{equation}
\H_N = \sqrt{\Lambda} \H_{N-1} + t_N \sum_{\s}
  \bigl( f_{N\s}^{\dag} f_{N-1,\s}^{\pdag} + \text{H.c.} \bigr)
  - E_{G,N},
\end{equation}
for $N = 1$, $2$, $3$, $\ldots$, where
$E_{G,N}$ is chosen so that the ground-state energy of $\H_N$ is zero.
$\H_N$ can be interpreted as describing a fermionic chain of length
$N+1$ sites with hopping coefficients that decay exponentially along
the chain away from the end (site $n=0$) to which the impurity and bosonic
degrees of freedom couple. The solution of $\H_N$ captures the dominant
physics at energies and temperatures of order $D\Lambda^{-N/2}$.

The NRG procedure is iterated until the problem reaches a fixed point at
which the spectrum of $\H_N$ and the matrix elements of all physical operators
between the eigenstates are identical to those of $\H_{N-2}$. (The
eigensolution of $\H_N$ differs from that of $\H_{N-1}$ even at a fixed point
due to odd-even alternation effects.\cite{Krishna-murthy:80})
In addition to the conduction-band discretization, two further approximations
must be imposed. First, the number of states on the fermionic chain grows by a
factor of 4 at each iteration, making it impractical to keep track of all
the many-body states beyond the first few iterations. Instead,
one retains just the $N_s$ many-particle states of lowest energy after
iteration $N$, creating a basis of dimension $4 N_s$ for iteration $N+1$.
Second, the presence of local bosons adds the further complication that the
full Fock space is infinite-dimensional even at iteration $N=0$, making
it necessary to restrict the maximum number of bosons to some finite number
$N_b$.

The NRG calculations reported in the sections that follow took advantage of the
conserved eigenvalues of the total spin-$z$ operator
\begin{equation}
\label{S_z}
\hat{S}_z = \frac{1}{2} \bigl( d_{\up}^{\dag} d_{\up}^{\pdag}
    - d_{\dn}^{\dag} d_{\dn}^{\pdag} \bigr)
    + \frac{1}{2} \sum_{n=0}^N \bigl( f_{n\up}^{\dag} f_{n\up}^{\pdag}
    - f_{n\dn}^{\dag} f_{n\dn}^{\pdag} \bigr) ,
\end{equation}
and the total ``charge'' operator
\begin{equation}
\label{Q}
\hat{Q} = \n_d - 1 + \sum_{n=0}^N
   \bigl( f_{n\up}^{\dag} f_{n\up}^{\pdag}
   + f_{n\dn}^{\dag} f_{n\dn}^{\pdag} - 1 \bigr)
\end{equation}
to reduce the Hamiltonian matrix to block-diagonal form, thereby reducing
the labor of matrix diagonalization.
In the absence of a magnetic field, the Hamiltonian $\H_N$ commutes not
only with $\hat{S}_z$, but also with the total spin raising and lowering
operators
\begin{equation}
\label{S+}
\hat{S}_+ = d_{\up}^{\dag} d_{\dn}^{\pdag}
    + \sum_n f_{n\up}^{\dag} f_{n\dn}^{\pdag}
\quad \text{and} \quad
\hat{S}_- = \bigl( \hat{S}_+ \bigr)^{\dag} ,
\end{equation}
the other two generators of SU(2) spin symmetry.
By analogy, one can interpret
\begin{equation}
\label{isospin}
\hat{I}_z = \frac{1}{2} \, \hat{Q}, \quad
\hat{I}_+ = - d_{\up}^{\dag} d_{\dn}^{\dag}
   + \sum_n (-1)^n f_{n\up}^{\dag} f_{n\dn}^{\dag}
\equiv \bigl( \hat{I}_- \bigr)^{\dag}
\end{equation}
as the generators of an SU(2) isospin symmetry.
Since $[\H_{\impboson}, \hat{I}_{\pm}]\not= 0$, $\H_N$ does not exhibit full
isospin rotation invariance, even though this symmetry turns out to be
recovered in the asymptotic low-energy behavior at each of the important
renormalization-group fixed points. In order to treat spin and charge
degrees of freedom on equal footing, we elected not to exploit total spin
conservation in our NRG calculations. However, in the sections that follow we
identify NRG states by their quantum number $S$ wherever appropriate.

Throughout the remainder of this paper, all energies are expressed as multiples
of the half-bandwidth $D=1$. Results are reported for the representative case
of a strongly correlated impurity level having $U = 0.5$ coupled to a local
bosonic mode of frequency $\omega_0 = 0.1$. The NRG calculations were performed
using discretization parameter $\Lambda = 2.5$ or 3, allowing up to $N_b = 40$
bosons, values found to yield well-converged results for the model parameters
considered. The number of retained many-body states was chosen sufficiently
large to eliminate discernible truncation errors in each computed quantity;
unless otherwise noted, this goal was attained using $N_s = 500$.

\section{Results: Particle-Hole-Symmetric Model With Band Exponent
$0 < r < \half$}
\label{sec:small-r_symm}

As reviewed in Sec.\ \ref{subsec:special-cases}, the particle-hole-symmetric
pseudogap Anderson model with a band exponent $0<r<\half$ has a QPT
at $\Gamma=\Gamma_c(r,U)$ between local-moment and symmetric
strong-coupling phases. In this section we investigate the changes that arise
from the Holstein coupling of the impurity charge to a local boson mode. For
bosonic couplings $\lambda<\lambda_0$ [see Eq.\ \eqref{lambda_0:def}], we find
that the low-energy physics of the pseudogap Anderson-Holstein model is largely
the same as for the pseudogap Anderson model with $U$ replaced by an effective
value $\Ueff$ [defined in Eq.\ \eqref{Ueff:def} below] that differs from
$\Ubar$ introduced in Sec.\ \ref{subsubsec:AHM}.
A QPT at $\Gamma=\Gamma_{c1}(r,U,\lambda<\lambda_0) \simeq\Gamma_c(r,\Ueff)$
exhibits universal properties indistinguishable from those at the critical
point of the pseudogap Anderson model. For stronger bosonic couplings
$\lambda > \lambda_0$, there is instead a QPT at
$\Gamma=\Gamma_{c2}(r,U,\lambda>\lambda_0)$ between the symmetric
strong-coupling phase and a local-charge phase in which the impurity has a
residual two-fold charge degree of freedom. The critical exponents describing
the local charge response at $\Gamma=\Gamma_{c2}$ are identical to those
characterizing the local spin response at $\Gamma=\Gamma_{c1}$.

All numerical results presented in this section were obtained in a zero or
infinitesimal magnetic field for a symmetric impurity with
$\Ed = -\half U = -0.25$, for a bosonic frequency $\omega_0 = 0.1$, and
for NRG discretization parameter $\Lambda = 3$.

\subsection{NRG spectrum and fixed points}
\label{subsec:small-r_symm:fixed_points}

The first evidence for the existence of multiple phases of the symmetric
pseudogap Anderson-Holstein model comes from the eigenspectrum of $\H_N$. This
spectrum can be used to identify stable and unstable renormalization-group fixed
points of the model.

\subsubsection{Weak bosonic coupling}

\begin{figure}[t]
\centering
\includegraphics[angle=270,width=0.95\columnwidth]{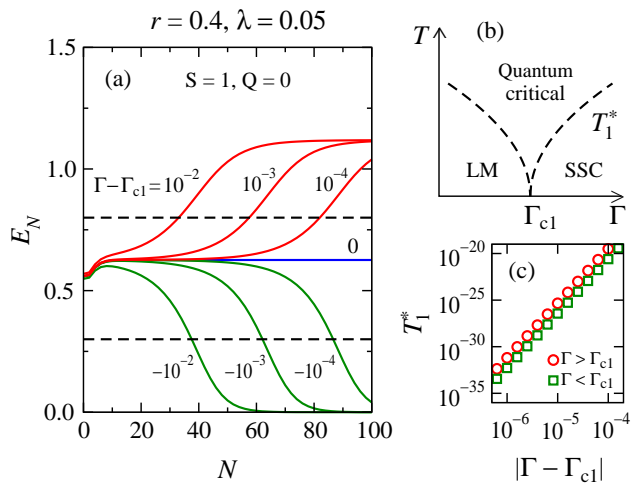}
\caption{\label{fig:small-r:NRGspec_s}
(Color online) Particle-hole-symmetric pseudogap Anderson-Holstein model near
its spin-sector critical point \Cs:
(a) NRG energy $E_N$ vs even iteration number $N$ of the first excited multiplet
having quantum numbers $S=1$, $Q=0$, calculated for $r=0.4$, $U=-2\Ed=0.5$,
$\omega_0 = 0.1$, $\lambda=0.05<\lambda_0\simeq 0.158$, and seven values of
$\Gamma-\Gamma_{c1}$ (with $\Gamma_{c1}\simeq 0.3166805$) labeled on the plot.
(b) Schematic phase diagram on the $\Gamma$--$T$ plane for $\lambda<\lambda_0$,
showing the $T=0$ transition between the local-moment ($\Gamma<\Gamma_{c1}$) and
symmetric strong-coupling  ($\Gamma>\Gamma_{c1}$) phases. Dashed lines mark the
scale $T^*_1$ of the crossover from the intermediate-temperature quantum-critical
regime to one or other of the stable phases.
(c) Crossover scale $T^*_1=D\Lambda^{-N^*_1/2}$ vs $|\Gamma-\Gamma_{c1}|$ in the
local-moment and symmetric strong-coupling phases, showing the power-law behavior
described in Eq.\ \eqref{nu1}. Here, $N^*_1$ is the interpolated value of $N$ at
which $E_N$ in (a) leaves its quantum-critical range by crossing one or other of
the horizontal dashed lines.}
\end{figure}

Figure \ref{fig:small-r:NRGspec_s}(a) shows---for $r=0.4$,
$\lambda=0.05<\lambda_0\simeq 0.158$, and seven different values of
$\Gamma$---the variation with even iteration number $N$ of the energy of the
first excited multiplet having quantum numbers $S=1$, $Q=0$.
For small values of $\Gamma$, this energy $E_N$ at first rises with
increasing $N$, but eventually falls towards the value $E_{\text{LM}}=0$
expected at the \textit{local-moment} fixed point corresponding to effective
model couplings $\Gamma=\lambda=0$ and $U=\infty$. At this fixed point, the
impurity $n_d=1$ doublet asymptotically decouples from the tight-binding chain
of length $N+1$, leaving a localized spin-$\half$ degree of freedom and
low-lying many-body excitations characterized by a Fermi-energy $s$-wave phase
shift $\delta_0(\epsilon=0)=0$, identical to that at the local-moment fixed
point of the pseudogap Anderson model (see Sec.\ \ref{subsubsec:AHM}).

For large $\Gamma$, $E_N$ instead rises monotonically to reach a limiting value
$E_{\text{SSC}}\simeq 1.119$ characteristic of the \textit{symmetric
strong-coupling} fixed point, corresponding to effective couplings
$\Gamma=\infty$ and $U=\lambda=0$. Here, the impurity level forms a spin singlet
with an electron on the end ($n=0$) site of the tight-binding chain. The singlet
formation ``freezes out'' the end site, leaving free-fermionic excitations on a
chain of reduced length $N$, leading to a Fermi-energy phase shift $\delta_0(0)=
-(1-r)(\pi/2)\,\sgn\,\veps$. This is the same phase shift as is found at the
symmetric strong-coupling fixed point of the pseudogap Anderson
model.\cite{Gonzalez-Buxton:98}

The local-moment and symmetric strong-coupling fixed points describe the
large-$N$ (low-energy $D\Lambda^{-N/2}$) physics for all initial choices of the
hybridization width except $\Gamma=\Gamma_{c1}\simeq 0.3166805$, in which special
case $E_N$ rapidly approaches $E_c\simeq 0.6258$ and remains at that energy up to
arbitrarily large $N$. This behavior can be associated with an unstable
\textit{critical point} \Cs\ separating the local-moment and symmetric
strong-coupling phases. (The subscript ``S'' indicates that \Cs\ separates
phases having different ground-state spin quantum numbers.) The critical point
corresponds to the pseudogap Anderson-Holstein model with $\lambda=0$ and
$\Gamma/U$ equaling some $r$-dependent critical value.

Whereas the critical coupling $\Gamma_{c1}$ is a nonuniversal function of all
the other model parameters ($r$, $U$, $\omega_0$, and $\lambda$), the low-energy
NRG spectra at the local-moment, symmetric strong-coupling, and \Cs\ fixed
points depend only on the band exponent $r$ and the NRG discretization parameter
$\Lambda$. For given $r$ and $\Lambda$, each spectrum is found to be identical
to that at the corresponding fixed point of the particle-hole-symmetric
pseudogap Anderson model. Not only can the spectrum be interpreted as arising
from an effective boson coupling $\lambda=0$, but it exhibits the SU(2) isospin
symmetry that is broken in the full pseudogap Anderson-Holstein model.

\subsubsection{Strong bosonic coupling}

\begin{figure}
\centering
\includegraphics[angle=270,width=0.95\columnwidth]{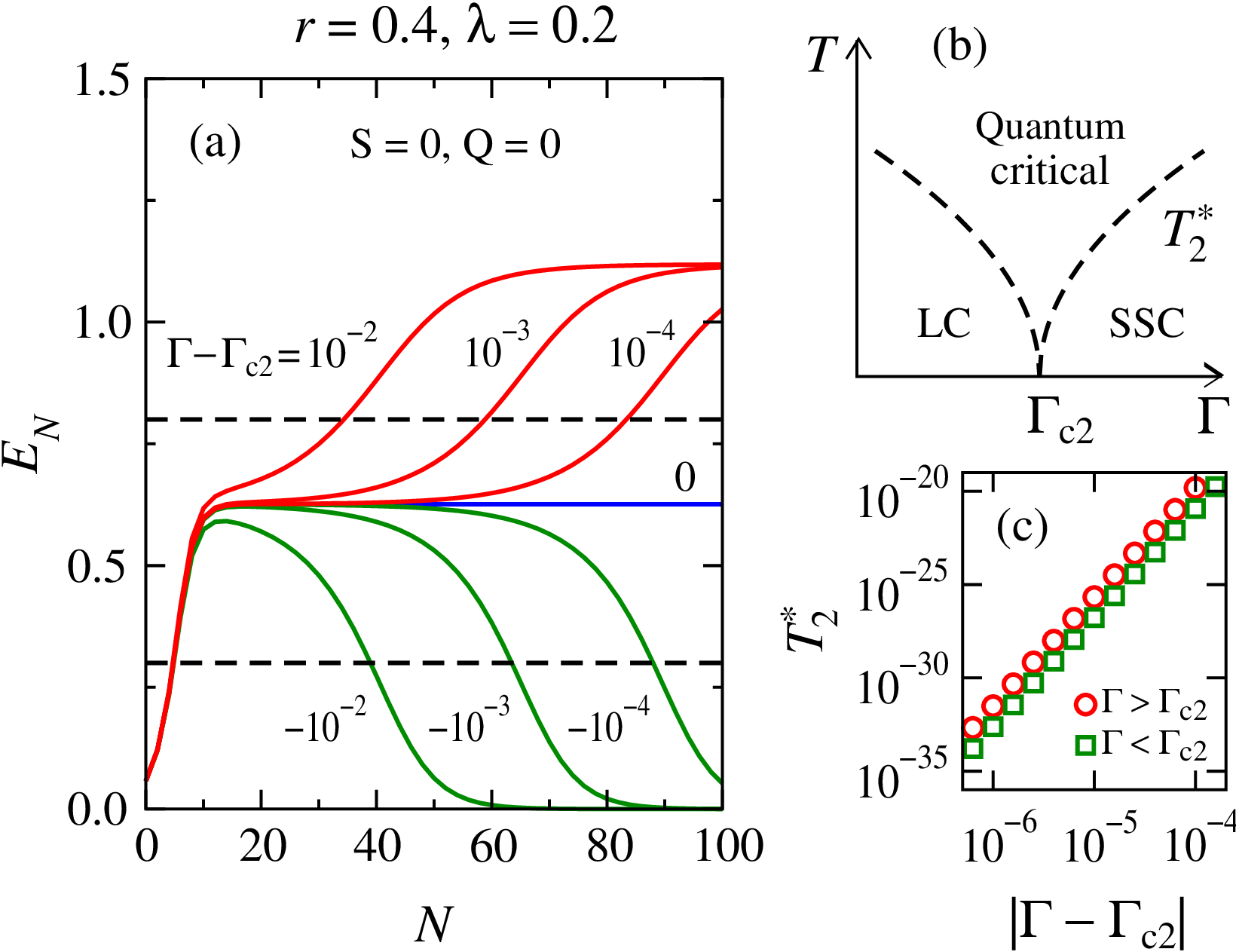}
\caption{\label{fig:small-r:NRGspec_c}
(Color online) Particle-hole-symmetric pseudogap Anderson-Holstein model near
its charge-sector critical point \Cc:
(a) NRG energy $E_N$ vs even iteration number $N$ of the first excited state
having quantum numbers $S=Q=0$, calculated for $r=0.4$, $U=-2\Ed=0.5$,
$\omega_0=0.1$, $\lambda=0.2>\lambda_0\simeq 0.158$, and seven values of
$\Gamma-\Gamma_{c2}$ (with $\Gamma_{c2}\simeq0.6878956$) labeled on the plot.
(b) Schematic $\Gamma$--$T$ phase diagram for $\lambda>\lambda_0$, showing the
$T=0$ transition between the local-charge and symmetric strong-coupling phases
and the scale $T^*_2$ of the crossover from the quantum-critical regime to a
stable phase.
(c) Crossover scale $T^*_2=D\Lambda^{-N^*_2/2}$ vs $|\Gamma-\Gamma_{c2}|$ in
the local-charge and symmetric strong-coupling phases, showing the power-law
behavior described in Eq.\ \eqref{nu2}. Here, $N^*_2$ is the interpolated value of
$N$ at which $E_N$ in (a) leaves its quantum-critical range by crossing one or
other of the horizontal dashed lines.}
\end{figure}

Figure \ref{fig:small-r:NRGspec_c}(a) plots the energy at even iterations of
the first NRG excited state having quantum numbers $S=Q=0$, for $r=0.4$,
$\lambda = 0.2 > \lambda_0 \simeq 0.158$, and seven different $\Gamma$ values.
For $\Gamma>\Gamma_{c2}\simeq0.6878956$, $E_N$ eventually flows to the value
$E_{\text{SSC}}\simeq 1.119$ identified in the weak-bosonic-coupling regime, and
examination of the full NRG spectrum confirms that the low-temperature behavior
is governed by the same symmetric strong-coupling fixed point.

For $\Gamma<\Gamma_{c2}$, $E_N$ flows to zero, the value found at the
local-moment fixed point. In fact, all the fixed-point many-body states obtained
for $\lambda>\lambda_0$ turn out to have the same energies as states at the
local-moment fixed point. However, the quantum numbers of states in the
$\lambda>\lambda_0$ spectrum and the local-moment spectra are not identical,
but rather are related by the interchanges $S\leftrightarrow I$ and
$S_z\leftrightarrow I_z$. We therefore associate the $\lambda>\lambda_0$
spectrum with a \textit{local-charge} fixed point, corresponding to
$\Gamma=\lambda=0$ and $U=-\infty$, at which the impurity has a residual
isospin-$\half$ degree of freedom. Like its local-moment counterpart, this
fixed point exhibits a phase shift $\delta_0(0) = 0$.

For $\Gamma=\Gamma_{c2}$, $E_N$ rapidly approaches and remains at the same
critical value $E_c$ as found for $\lambda<\lambda_0$ and $\Gamma=\Gamma_{c1}$.
Once again, however, the many-body spectrum is related to that at the
corresponding weak-bosonic-coupling fixed point by interchange of spin and
isospin quantum numbers, leading to the interpretation of this fixed point as
a charge analog \Cc\ of the critical point of the particle-hole-symmetric
pseudogap Anderson model.

\subsection{Phase boundaries}
\label{subsec:small-r_symm:boundaries}

\begin{figure}[t]
\centering
\includegraphics[angle=270,width=0.85\columnwidth]{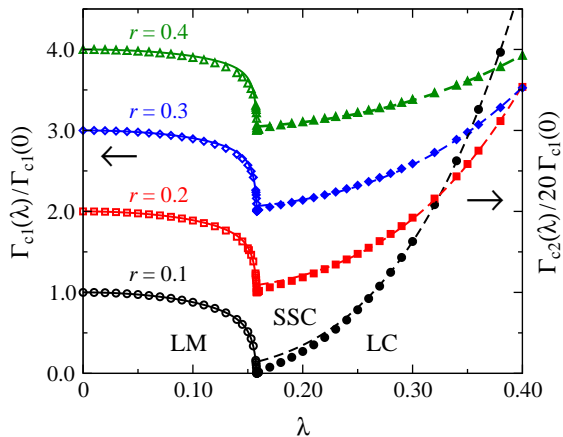}
\caption{\label{fig:small-r:phase_bound}
(Color online) Phase boundaries of the particle-hole-symmetric pseudogap
Anderson-Holstein model in zero magnetic field: Variation with bosonic coupling
$\lambda$ of the critical hybridization widths $\Gamma_{c1}$ and $\Gamma_{c2}$.
Data for $U=-2\Ed=0.5$, $\omega_0=0.1$, and band exponents $r=0.1$, 0.2,
0.3, and 0.4 are scaled and offset for clarity: the quantities plotted are
$\Gamma_{c1}(\lambda)/\Gamma_{c1}(0)+10r-1$ (empty symbols) and
$\Gamma_{c2}(\lambda)/20\Gamma_{c1}(0)+10r-1$ (filled symbols). Solid lines
show the prediction of Eq.\ \protect\eqref{Gamma_c1_vs_lambda}, while dashed
lines plot the form $A(r) \lambda^{2(2-r)}$ suggested by Eq.\
\eqref{Gamma_c2_vs_lambda} with a prefactor $A(r)$ determined by fitting
values of $\Gamma_{c2}$ for $0.3\le \lambda \le 0.4$.}
\end{figure}

Figure \ref{fig:small-r:phase_bound} shows phase boundaries for the symmetric
pseudogap Anderson-Holstein model, as established for $U=-2\Ed=0.5$, and
$\omega_0=0.1$ by examination of the NRG spectrum. In the atomic limit
$\Gamma=0$ we find a level-crossing transition between the local-moment and
local-charge phases at $\lambda=\lambda_0\simeq 0.15812(1)$, a value in
excellent agreement with the prediction of Eq.\ \eqref{lambda_0:def}.
(Throughout this paper, a digit in parentheses following a number indicates the
estimated nonsystematic error in the last digit of the number.)
For each of four values of the band exponent $r$, the figure plots the
critical hybridization widths $\Gamma_{c1}$ (open symbols,
for $0\le\lambda<\lambda_0$) and $\Gamma_{c2}$ (filled symbols, for
$\lambda>\lambda_0$) normalized by the $\lambda=0$ value of $\Gamma_{c1}$,
which coincides with the critical hybridization width $\Gamma_c$ of the
corresponding pseudogap Anderson model. The lines represent analytical
expressions for the phase boundaries that will be explained in the remainder
of this subsection.

In the particle-hole-symmetric pseudogap Anderson model, the critical
hybridization width $\Gamma_c(r,U)$ can be established by performing a
Schrieffer-Wolff transformation\cite{Schrieffer:66} that maps the problem to
a pseudogap Kondo model with a Kondo exchange coupling
\begin{equation}
\label{SW:PAM}
\rho_0 J = \frac{8\Gamma}{\pi U} \left( \frac{U}{2D} \right)^r .
\end{equation}
Here, the Kondo coupling in a conventional metal ($r=0$) is multiplied by a
factor $(U/2D)^r$ that accounts for the irrelevance of the hybridization width
under poor-man's scaling [see Eq.\ \eqref{Gamma:scaling}] while neglecting the
much weaker renormalization of the on-site interaction [Eq.\ \eqref{U:scaling}].
The critical coupling $J_c$ of the particle-hole-symmetric pseudogap Kondo model
satisfies $\rho_0 J_c = j_c(r)$, where $j_c(r)\simeq r$ for $r\ll\half$ (Ref.\
\onlinecite{Withoff:90}) and $j_c(r)\to\infty$ for $r\to\half$
(Refs.\ \onlinecite{Ingersent:96}, \onlinecite{Gonzalez-Buxton:98}, and
\onlinecite{Logan:00}). Combining this condition with Eq.\ \eqref{SW:PAM} yields
\begin{equation}
\label{Gamma_c:PAM}
\Gamma_c = \frac{\pi}{4} j_c(r) \, D \left( \frac{U}{2D} \right)^{1-r} .
\end{equation}
It is important to note that an equivalent expression has been derived within the
local-moment approach to the pseudogap Anderson model without reference to a
Schrieffer-Wolff transformation [see Eq.\ (6.10b) of Ref.\ \onlinecite{Logan:00}]
and has been verified via NRG calculations.\cite{Bulla:00} As such, Eq.\
\eqref{Gamma_c:PAM} with a suitably chosen value of $j_c(r)$ is applicable even
for $r$ approaching $\frac{1}{2}$ where charge fluctuations for
$\Gamma\simeq\Gamma_c$ are too strong to allow mapping to a Kondo
model.\cite{Gonzalez-Buxton:98} We now consider how Eq.\ \eqref{Gamma_c:PAM}
should be modified to describe the phase boundaries of the pseudogap
Anderson-Holstein model.

\subsubsection{Weak bosonic coupling}

For $\lambda\ll\lambda_0$ (and hence $\Ubar>0$), it has been
shown\cite{Cornaglia:04} that a generalized Schrieffer-Wolff transformation
maps the particle-hole-symmetric Anderson-Holstein model to a Kondo model with
a dimensionless exchange coupling
\begin{equation}
\label{rho J:weak}
\rho_0 J = \frac{4\Gamma}{\pi} \: e^{-(\lambda/\omega_0)^2}
   \sum_{\nb = 0}^{\infty} \frac{1}{\nb!} \,
   \frac{(\lambda/\omega_0)^{2\nb}}{\Ubar/2 + \nb\,\omega_0} ,
\end{equation}
representing a sum over virtual transitions of the impurity from occupation
$n_d=1$ to $n_d=0$ or $2$, accompanied by excitation of different numbers
$\nb=0$, $1$, $\ldots$ of bosonic quanta. To facilitate comparison with the
corresponding expression $\rho_0 J=8\Gamma/\pi U$ for the symmetric Anderson
model without bosons, one can use Eq.\ \eqref{S:def} to recast Eq.\
\eqref{rho J:weak} in the form
\begin{align}
\label{J_SW}
\rho_0 J
&= \frac{8\Gamma}{\pi|\Ubar|} \: S\biggl(\frac{|\Ubar|}{2\omega_0}, \:
   \frac{\Ep}{\omega_0}\biggr) \notag \\
&\equiv \frac{8\Gamma}{\pi\Ueff} \, ,
\end{align}
where, both for $\Ubar>0$ (as is the case here) and for $\Ubar<0$,
\begin{equation}
\label{Ueff:def}
\Ueff = 2 \, \E(|\Ubar|/2)
\end{equation}
with $\E(E)$ as defined in Eq.\ \eqref{E:def}. Equation \eqref{J_SW} suggests
that $\Ueff$ plays the role of an effective Coulomb repulsion in the
low-energy many-body physics of the full Anderson-Holstein model, distinct from
the quantity $\Ubar$ [Eq.\ \eqref{Ubar:def}] that emerges from considering just
the atomic limit $\Gamma=0$. Like $\Ubar$, $\Ueff$ passes through zero at
$\lambda=\lambda_0$. For fixed $U$, the ratio $\Ueff/\Ubar$ evolves smoothly
from $1$ for $\lambda=0$ to $e^{U/2\omega_0}$ ($\simeq 12$ in the case
$U/2\omega_0 = 2.5$ used in our calculations) for $\lambda=\lambda_0$ to $2$
for $\lambda\to\infty$.

Extension of the analysis of Ref.\ \onlinecite{Cornaglia:04} to the
case of a pseudogap density of states leads to the conclusion that the
critical hybridization width separating the local-moment and symmetric
strong-coupling phases should satisfy
\begin{equation}
\label{Gamma_c1_vs_lambda}
\Gamma_c = \frac{\pi}{4} j_c(r) \, D \left( \frac{\Ueff}{2D} \right)^{1-r} .
\end{equation}
The solid lines plotted in Fig.\ \ref{fig:small-r:phase_bound} show the
boundaries predicted by Eq.\ \eqref{Gamma_c1_vs_lambda} with numerical
evaluation of $\Ueff$. The agreement with the NRG data points is excellent
for all four values of $r$, and for $\lambda$ extending from zero almost all
the way to $\lambda_0$.

\subsubsection{Strong bosonic coupling}

Cornaglia \emph{et al}.\ have demonstrated\cite{Cornaglia:04} that the
Anderson-Holstein model with $\lambda\gg\lambda_0$ maps to a charge analog of
the Kondo model in which the impurity isospin degree of freedom [the impurity
($d$-electron) parts of the operators defined in Eqs.\ \eqref{isospin}]
is screened by its conduction-band counterpart. The impurity-band
isospin exchange is anisotropic, with a longitudinal coupling
$\rho_0 J_z = 8\Gamma / \pi |\Ueff|$ and a transverse coupling
\begin{equation}
\rho_0 J_{\p}
= \frac{8\Gamma}{\pi|\Ubar|} \: S\biggl(\frac{|\Ubar|}{2\omega_0}, \:
  -\frac{\Ep}{\omega_0}\biggr)
\sim e^{-2\Ep/\omega_0} \, \rho_0 J_z ,
\end{equation}
where $\Ueff$ and $S(a,x)$ are defined in Eqs.\ \eqref{Ueff:def} and
\eqref{S:def}, respectively.
Closer investigation shows that an approximation that is equivalent
for large $\lambda$ but also remains valid much closer to $\lambda_0$ is
\begin{equation}
\label{aniso:ratio}
J_{\p} \simeq e^{-|\Ueff|/2\omega_0} J_z .
\end{equation}
The strong suppression of ``charge-flip'' scattering arises from the
exponentially small overlap between the ground state of the displaced harmonic
oscillator that minimizes the electron-boson interaction in the sector $n_d=0$
and the corresponding ground state for $n_d=2$ (see Sec.\ \ref{subsubsec:AHM}).

A poor-man's scaling analysis of the anisotropic pseudogap Kondo
model\cite{Cheng:inprep} indicates that for $J_z \gg |J_{\p}|$, the phase
boundary is defined by a condition $\rho_0 J_z \simeq r \ln(2 J_z/J_{\p})$.
Applying this condition to the pseudogap Anderson-Holstein model, carrying over
Eq.\ \eqref{aniso:ratio} from the case $r=0$, and assuming [by analogy with Eq.\
\eqref{SW:PAM}] that $\rho_0 J_z \propto |\Ueff/D|^{r-1}$, yields
\begin{equation}
\label{Gamma_c2_vs_lambda}
\Gamma_{c2}(\lambda) \sim \left| \frac{\Ueff}{D} \right|^{2-r}
  \sim \lambda^{2(2-r)},
\end{equation}
where we have used $\Ueff \simeq 2 \Ubar \simeq -4\,\Ep = -4\lambda^2/\omega_0$
for $\lambda \gg \lambda_0$. The validity of Eq.\ \eqref{Gamma_c2_vs_lambda} is
questionable because the critical hybridization widths it demands are too large
to justify mapping to a Kondo model. Nonetheless, the NRG data for each value
of $r$ plotted in Fig.\ \ref{fig:small-r:phase_bound} follow a
$\lambda^{2(2-r)}$ dependence (dashed lines) quite closely for
$0.2\le\lambda\le 0.4$. (This power law must break down closer to the level
crossing between the local-moment and local-charge phases because $\Gamma_{c2}$
necessarily vanishes at $\lambda=\lambda_0$.)

\subsection{Crossover scales}
\label{subsec:small-r_symm:crossovers}

Aside from allowing the identification of renormalization-group fixed points and
phase boundaries, the eigenspectrum of $\H_N$ can also be used to define
temperature scales characterizing crossovers between the domains of influence
of different fixed points. We focus on the smallest such scale, which describes
the approach to one of the stable fixed points of the problem.

\subsubsection{Weak bosonic coupling}

With decreasing $|\Gamma-\Gamma_{c1}|$, $E_N$ in Fig.\
\ref{fig:small-r:NRGspec_s}(a) remains close to its critical value $E_c$ over
an increasing number of iterations before heading either to $E_{\text{LM}}$ or
to $E_{\text{SSC}}$. To quantify this effect, it is useful to define threshold
energy values $E_{\pm}$ where $E_{\text{LM}}<E_-<E_c<E_+<E_{\text{SSC}}$. The
passage of $E_N$ below $E_-$ (above $E_+$) at some $N^*_1$---determined by
interpolation of the NRG data at even integer values of $N$---can be taken to
mark the crossover around temperature $T_1^* \simeq D\Lambda^{-N^*_1/2}$ from
an intermediate-temperature quantum-critical regime dominated by the unstable
critical point \Cs\ to a low-temperature regime controlled by the stable
local-moment (symmetric strong-coupling) fixed point. This crossover scale is
expected to vanish for $\Gamma\to\Gamma_{c1}$, as shown schematically in Fig.\
\ref{fig:small-r:NRGspec_s}(b).
Figure \ref{fig:small-r:NRGspec_s}(c) plots values of $T^*_1$ determined by the
criterion $E_-=0.3$, $E_+=0.8$. These data are consistent with the relation
\begin{equation}
\label{nu1}
T^*_1 \propto |\Gamma-\Gamma_{c1}|^{\nu_1} \quad \text{as} \quad \Gamma
\to \Gamma_{c1},
\end{equation}
where $\nu_1$ is the correlation-length exponent at the quantum critical point.
The numerical value of $\nu_1$ is independent of the precise choice of the
thresholds $E_{\pm}$. What is more, different combinations of the model
parameters $r$, $U$, $\omega_0$, and $\lambda$ result in different critical
couplings $\Gamma_{c1}$, but $\nu_1$ depends only on the band exponent $r$.
Values for three representative cases are listed in Table \ref{tab:small-r}.

\begin{table}
\caption{\label{tab:small-r}
Critical exponents at the critical point \Cs\ of the particle-hole-symmetric
pseudogap Anderson-Holstein model, evaluated for three band exponents $r$.
The critical exponents are defined in Eqs.\ \protect\eqref{nu1} and
\protect\eqref{exponents1}. A number in parentheses indicates the estimated
random error in the last digit of each exponent.}
\begin{tabular*}{\columnwidth}{@{\extracolsep{\fill}}llllll}
\hline\hline \\[-2ex]
\multicolumn{1}{c}{$r$} & \multicolumn{1}{c}{$\nu_1$} &
  \multicolumn{1}{c}{$\beta_1$}& \multicolumn{1}{c}{$1/\delta_1$} &
  \multicolumn{1}{c}{$x_1$} & \multicolumn{1}{c}{$\gamma_1$} \\[.2ex]
\hline \\[-2ex]
0.2 & 6.22(1) & 0.15(1)  & 0.02630(2)  & 0.9488(2) & 5.85(6) \\
0.3 & 5.14(1) & 0.34(1)  & 0.07364(1)  & 0.8629(3) & 4.41(3) \\
0.4 & 5.84(1) & 0.90(1)  & 0.1845(1)   & 0.6885(2) & 3.95(5) \\[.2ex]
\hline\hline
\end{tabular*}
\end{table}

\subsubsection{Strong bosonic coupling}

The passage of $E_N$ outside a range $E_-<E_N<E_+$ can also be used to define a
crossover scale near its charge-sector critical point. This scale $T^*_2$ is
expected to vanish at the critical point according to
\begin{equation}
\label{nu2}
T^*_2 \propto |\Gamma-\Gamma_{c2}|^{\nu_2} \quad \text{as} \quad \Gamma
\to \Gamma_{c2},
\end{equation}
a behavior that is sketched qualitatively in Fig.\
\ref{fig:small-r:NRGspec_c}(b)
and is confirmed quantitatively in Fig.\ \ref{fig:small-r:NRGspec_c}(c).
For all the values of $r$ and $\Lambda$ that we have studied, the numerical
values of $\nu_1$ and $\nu_2$ coincide to within our estimated errors.

\subsection{Impurity thermodynamic properties}
\label{subsec:small-r_symm:imp_props}

This section addresses the variation with temperature of the impurity
contributions\cite{impurity-props} to the static spin and charge
susceptibilities and to the entropy.
With the conventional definitions $T\chi_{s,\imp}=\langle\hat{S}_z^2\rangle$
and $T\chi_{c,\imp}=\langle Q^2\rangle$, a symmetric impurity level isolated
from the conduction band ($\Gamma=0$) has $T\chi_{s,\imp}=1/4$ and
$T\chi_{c,\imp}=0$ for $\Ueff\gg T$, but $T\chi_{c,\imp}=1$ and
$T\chi_{s,\imp}=0$ for $\Ueff\ll -T$, with $\Ueff$ as defined in
Eq.\ \eqref{Ueff:def}. Due to the factor of 4 difference between the
local-moment spin susceptibility and the charge susceptibility of a local
charge doublet, it is most appropriate to compare $T\chi_{s,\imp}$ with
$\frac{1}{4}T\chi_{c,\imp}$. During the NRG calculation of these thermodynamic
properties, $N_s = 3\,000$ states were retained after each iteration.

\subsubsection{Weak bosonic coupling}

\begin{figure}
\centering
\includegraphics[angle=270,width=0.95\columnwidth]{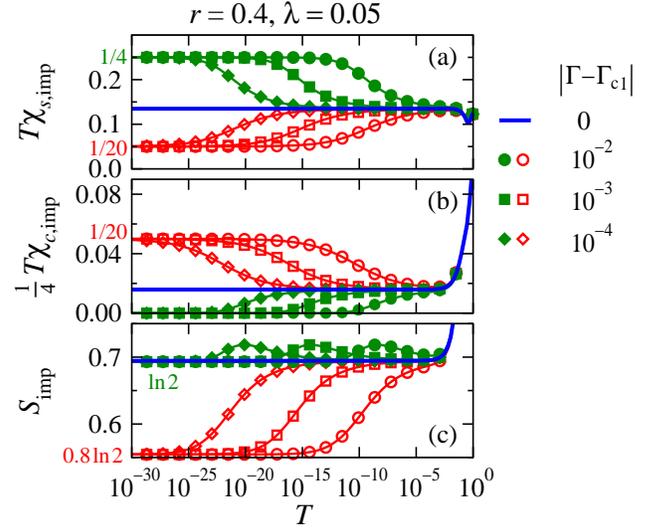}
\caption{\label{fig:r=0.4_symm:thermo_s}
(Color online) Thermodynamic properties of the particle-hole-symmetric
pseudogap Anderson-Holstein model near its spin-sector critical point \Cs:
Temperature dependence of the impurity contribution to
(a) the static spin susceptibility $\chi_{s,\imp}$ multiplied by temperature,
(b) the static charge susceptibility $\chi_{c,\imp}$ multiplied by temperature,
and (c) the entropy $S_{\imp}$,
for $r=0.4$, $U=-2\Ed=0.5$, $\omega_0 = 0.1$,
$\lambda=0.05<\lambda_0\simeq 0.158$, and the seven values of
$\Gamma-\Gamma_{c1}$ labeled in the legend. Filled (open) symbols connected
by guiding lines represent data in the local-moment (symmetric strong-coupling)
phase, while thick lines without symbols show the critical properties at \Cs.
$N_s = 3\,000$ states were retained after each NRG iteration.}
\end{figure}

Figure \ref{fig:r=0.4_symm:thermo_s} plots the temperature dependence of
$T\chi_{s,\imp}$, $\frac{1}{4}T\chi_{c,\imp}$, and $S_{\imp}$ for $r=0.4$,
$\lambda=0.05$, and seven values of $\Gamma$ straddling $\Gamma_{c1}$. At high
temperatures $T\gg\max(\Ueff,\Gamma)$, the properties lie close to those of the
free-orbital fixed point ($T\chi_{s,\imp}=\frac{1}{4}T\chi_{c,\imp}=1/8$ and
$S_{\imp}=\ln 4$), irrespective of the specific value of $\Gamma$. However, the
$T\to 0$ behaviors directly reflect the existence of a QPT
at $\Gamma=\Gamma_{c1}$. In the local-moment phase ($\Gamma<\Gamma_{c1})$, the
residual impurity spin doublet is revealed in the limiting behaviors
$T\chi_{s,\imp}=1/4$, $\frac{1}{4}T\chi_{c,\imp}=0$, and $S_{\imp}=\ln 2$. In
the symmetric strong-coupling phase ($\Gamma>\Gamma_{c1})$, the impurity
degrees of freedom are quenched to the maximum extent possible given the
power-law hybridization function,\cite{Gonzalez-Buxton:98} yielding
$T\chi_{s,\imp}=\frac{1}{4}T\chi_{c,\imp}=r/8$ and
$S_{\imp}=2r\ln 2$. Exactly at $\Gamma=\Gamma_{c1}$, the low-temperature
properties are distinct from those in either phase: $T\chi_{s,\imp}\simeq
0.1348$, $\frac{1}{4}T\chi_{c,\imp}\simeq 0.0158$, and $S_{\imp}\simeq 0.694
\simeq\ln 2$. These values vary with the band exponent $r$, but are independent
of other model parameters such as $U$, $\omega_0$, and $\lambda$, so they can
be regarded as characterizing the critical point \Cs. For all the $r$ values
that we have examined, the critical properties coincide with those at the
corresponding critical point of the pseudogap Kondo or Anderson
models.\cite{Gonzalez-Buxton:98}

When $\Gamma$ deviates slightly from $\Gamma_{c1}$, the thermodynamic properties
follow their critical behaviors at high temperatures, but cross over for
$T\lesssim T^*_1$ to approach the values characterizing the local-moment or
symmetric strong-coupling phase. The crossover temperature $T^*_1$ coincides up
to a constant multiplicative factor with that extracted from the NRG spectrum
(as described in Sec.\ \ref{subsec:small-r_symm:boundaries}) and its variation
with $|\Gamma-\Gamma_{c1}|$ yields, via Eq.\ \eqref{nu1}, a correlation-length
exponent $\nu_1(r)$ in agreement with the values listed in Table
\ref{tab:small-r}.

\subsubsection{Strong bosonic coupling}

\begin{figure}
\centering
\includegraphics[angle=270,width=0.95\columnwidth]{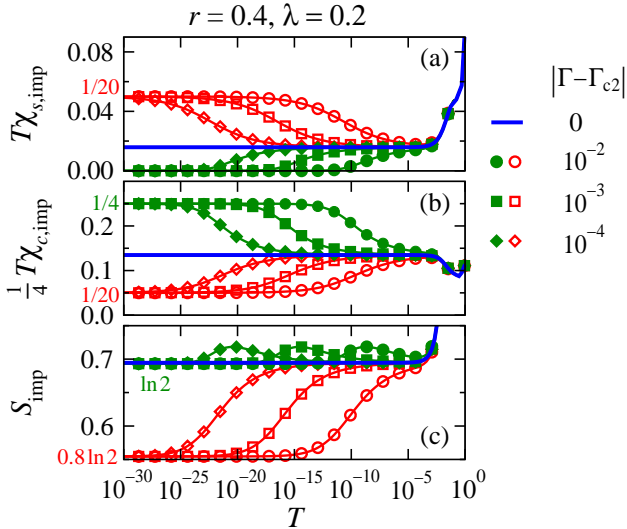}
\caption{\label{fig:r=0.4_symm:thermo_c}
(Color online) Thermodynamic properties of the particle-hole-symmetric
pseudogap Anderson-Holstein model near its charge-sector critical point \Cc:
Temperature dependence of the impurity contribution to
(a) the static spin susceptibility $\chi_{s,\imp}$ multiplied by temperature,
(b) the static charge susceptibility $T\chi_{c,\imp}$ multiplied by
temperature, and (c) the entropy $S_{\imp}$, for $r=0.4$, $U=-2\Ed=0.5$,
$\omega_0 = 0.1$, $\lambda=0.2>\lambda_0\simeq 0.158$, and the seven values of
$\Gamma-\Gamma_{c2}$ labeled in the legend. Filled (open) symbols connected
by guiding lines represent data in the local-charge (symmetric strong-coupling)
phase, while thick lines without symbols show the critical properties at \Cc.
$N_s = 3\,000$ states were retained after each NRG iteration.}
\end{figure}

Figure \ref{fig:r=0.4_symm:thermo_c} plots the temperature dependence of
$T\chi_{s,\imp}$, $\frac{1}{4}T\chi_{c,\imp}$, and $S_{\imp}$ for $r=0.4$,
$\lambda=0.2$, and various $\Gamma$ straddling the critical value
$\Gamma_{c2}$. Like in the case of weak bosonic coupling, the $T\to 0$
behaviors distinguish the two stable phases: the properties $T\chi_{s,\imp}=0$,
$\frac{1}{4}T\chi_{c,\imp}=1/4$, and $S_{\imp}=\ln 2$ in the local-charge phase
contrast with $T\chi_{s,\imp}=\frac{1}{4}T\chi_{c,\imp}=r/8$ and
$S_{\imp}=2r\ln 2$ in the symmetric strong-coupling phase. Exactly at
$\Gamma=\Gamma_{c2}$, $T\chi_{s,\imp}\simeq 0.0158$,
$\frac{1}{4}T\chi_{c,\imp}\simeq 0.1348$, and $S_{\imp}\simeq 0.694$, values
that can be taken to characterize the critical point \Cc. From the
thermodynamics near \Cc, one can extract a crossover scale $T^*_2$
that gives [via Eq.\ \eqref{nu2}] a correlation-length exponent $\nu_2$
identical to that determined from the NRG spectrum.

Figures \ref{fig:r=0.4_symm:thermo_s} and \ref{fig:r=0.4_symm:thermo_c}
illustrate the general property that the temperature dependence of the spin
(charge) susceptibility at \Cs\ mirrors that of the charge (spin) susceptibility
at \Cc, while the entropy behaves in the same manner at both critical points.
These observations are consistent with the equivalence of the NRG spectra at the
two fixed points under interchange of spin and charge quantum numbers (see
Sec.\ \ref{subsec:small-r_symm:fixed_points}).

\subsection{Local response and universality class}
\label{subsec:small-r_symm:loc_props}

In order to investigate in greater detail the properties of the spin and charge
critical points (\Cs\ and \Cc\ in Fig.\ \ref{fig:small-r:RG_flow}), it is
necessary to identify an appropriate order parameter for each QPT.
The symmetric strong-coupling and local-moment phases can be
distinguished by their values (0 and $\half$, respectively) of the magnitude
$|\langle S_z\rangle|$ of the total spin in a vanishingly small magnetic
field applied along the $z$ direction. Similarly, the magnitude
$|\langle Q\rangle|$ of the total charge in the presence of an infinitesimal
electric potential takes the value 0 in the symmetric strong-coupling phase
and $1$ in the local-charge phase. However, the fact that $S_z$ and $Q$ are
conserved quantities---i.e., that the pseudogap Anderson-Holstein
Hamiltonian commutes with $\hat{S}_z$ and $\hat{Q}$ defined in
Eqs.\ \eqref{S_z} and  \eqref{Q}, respectively---prevents these candidate
order parameters from exhibiting nontrivial critical
exponents.\cite{Sachdev:94,Vojta:01} Instead,
we must look to the impurity response to local fields in order to probe the
quantum critical behavior.

\subsubsection{Weak bosonic coupling}

In the pseudogap Kondo and Anderson models, the critical properties manifest
themselves\cite{Ingersent:02} through the response to a local magnetic field
$h$ that couples only to the impurity spin as specified in Eq.\ \eqref{h:def}.
The order parameter for the pseudogap QPT is the limiting value as $h\to 0$
of the local moment
\begin{equation}
M_{\loc}=\langle \half(\n_{d\up}-\n_{d\dn})\rangle,
\end{equation}
and the order-parameter susceptibility is the static local spin susceptibility
\begin{equation}
\chi_{s,\loc}=-\lim_{h\to0}\frac{M_{\loc}}{h}.
\end{equation}

Based on the similarities noted above between the pseudogap-Anderson critical
point and the \Cs\ critical point of the pseudogap Anderson-Holstein model
(i.e., the properties of the phases on either side of each transition, the NRG
spectrum at the transition, and the value of the order-parameter exponent), we
expect that the two QPTs also to share the same order
parameter. Accordingly, the behaviors of $M_{\loc}$ and $\chi_{s,\loc}$ in the
vicinity of the critical hybridization width $\Gamma=\Gamma_{c1}$ should be
described by critical exponents $\beta_1$, $\gamma_1$, $\delta_1$, and $x_1$
defined as follows:
\begin{subequations}
\label{exponents1}
\begin{align}
M_{\loc}(\Gamma<\Gamma_{c1}; h\to0, T=0)&
  \propto (\Gamma_{c1}-\Gamma)^{\beta_1}, \\
\chi_{s,\loc}(\Gamma>\Gamma_{c1}; T=0)&
  \propto (\Gamma-\Gamma_{c1})^{-\gamma_1}, \\
M_{\loc}(h; \Gamma=\Gamma_{c1}, T=0)& \propto |h|^{1/\delta_1}, \\
\chi_{s,\loc}(T; \Gamma=\Gamma_{c1})& \propto T^{-x_1}.
\end{align}
\end{subequations}

\begin{figure}[t]
\centering
\includegraphics[angle=270,width=\columnwidth]{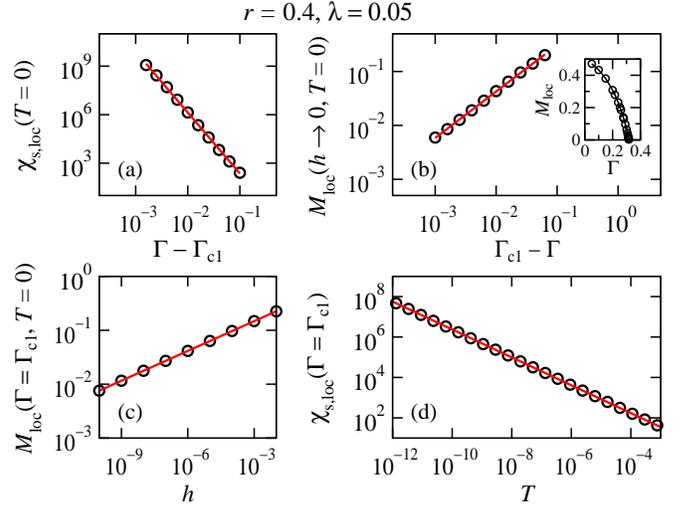}
\caption{\label{fig:small-r:local_s}
(Color online) Static local spin response of the particle-hole-symmetric
pseudogap Anderson-Holstein model near its spin-sector critical point \Cs.
Circles are NRG data for $r=0.4$, $U=-2\Ed=0.5$, $\omega_0=0.1$, and
$\lambda=0.05$, at or near the critical hybridization width
$\Gamma_{c1}\simeq 0.3166805$. Straight lines represent power-law fits.
(a) Static local spin susceptibility $\chi_{s,\loc}(T=0)$
vs $\Gamma-\Gamma_{c1}$ in the symmetric strong-coupling phase.
(b) Local magnetization $M_{\loc}(h\to0, T=0)$ vs
$\Gamma_{c1}-\Gamma$ in the local-moment phase. Inset: Continuous vanishing of
$M_{\loc}(h\to0, T=0)$ as $\Gamma$ approaches $\Gamma_{c1}$ from below.
(c) Local magnetization $M_{\loc}(\Gamma=\Gamma_{c1}, T=0)$ vs local
magnetic field $h$.
(d) Static local spin susceptibility
$\chi_{s,\loc}(\Gamma=\Gamma_c)$ vs temperature $T$.}
\end{figure}

The preceding expectations are proved correct by NRG calculations, as
demonstrated in Fig.\ \ref{fig:small-r:local_s} for $r=0.4$ and $\lambda=0.05$,
the case treated in Fig.\ \ref{fig:small-r:NRGspec_s}. The critical exponents
extracted as best-fit slopes of log-log plots are listed in Table
\ref{tab:small-r} for three values of the band exponent $r$. The values of
individual critical exponents vary with $r$, but are independent of other
Hamiltonian parameters ($U$, $\omega_0$, and $\lambda$) and are well converged
with respect to the NRG parameters $(\Lambda$, $N_s$, and $N_b$). To within
their estimated accuracy, the critical exponents for a given $r$ obey the
hyperscaling relations
\begin{equation}
\label{hyper}
\delta_1=\frac{1+x_1}{1-x_1} , \quad 2\beta_1=\nu(1-x_1) ,
\quad \gamma_1=\nu_1 x_1 ,
\end{equation}
which are consistent with the scaling ansatz
\begin{equation}
\label{ansatz}
F=T \, \mathcal{F} \left( \frac{|\Gamma-\Gamma_{c1}|}{T^{1/\nu_1}} \, ,
\frac{|h|}{T^{(1+x_1)/2}} \right)
\end{equation}
for the nonanalytic part of the free energy at an interacting critical
point.\cite{Ingersent:02}

\subsubsection{Strong bosonic coupling}

\begin{figure}[t]
\centering
\includegraphics[angle=270,width=\columnwidth]{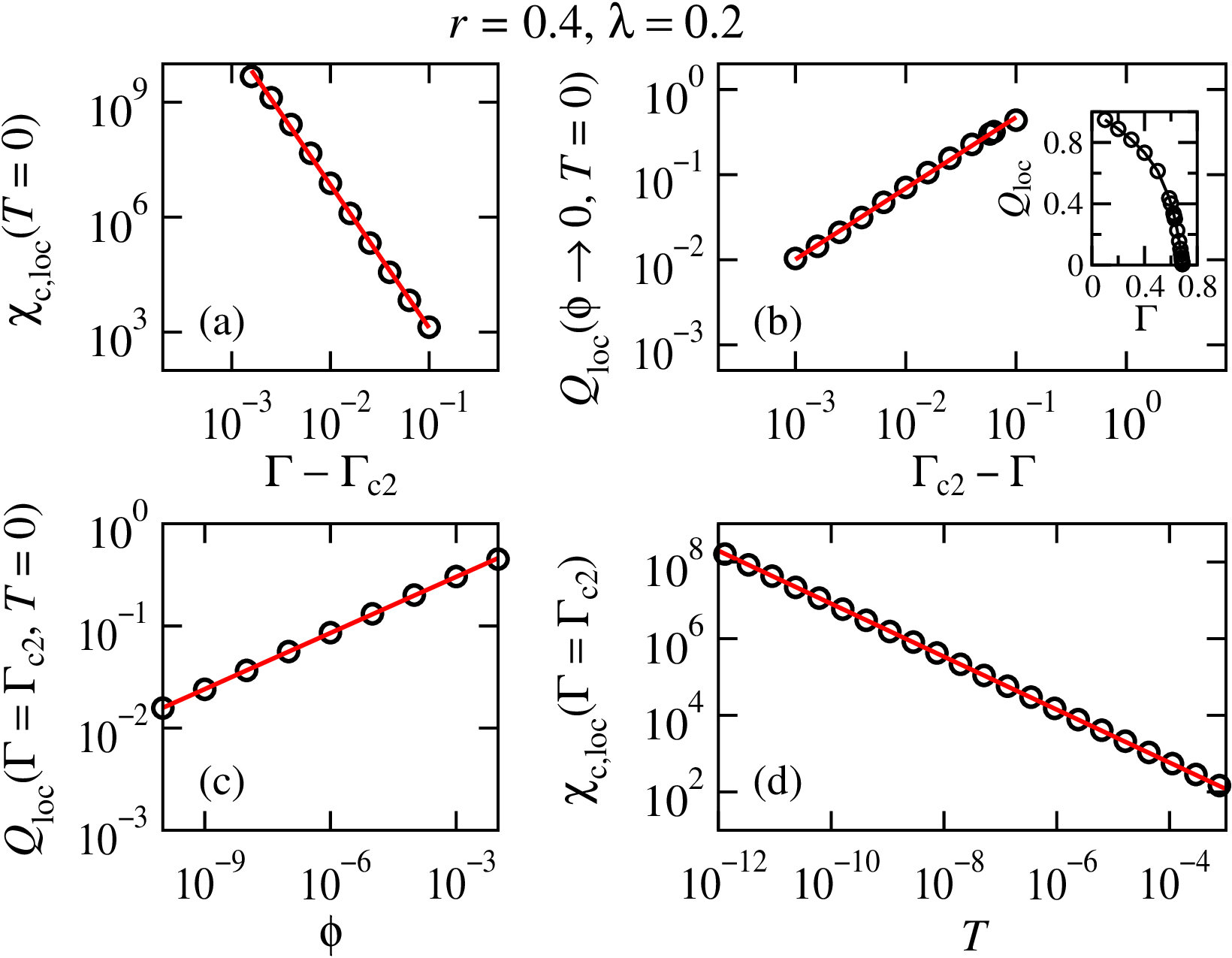}
\caption{\label{fig:small-r:local_c}
(Color online) Static local charge response of the particle-hole-symmetric
pseudogap Anderson-Holstein model near its charge-sector critical point \Cc.
Circles are NRG data for $r=0.4$, $U=-2\Ed=0.5$, $\omega_0=0.1$, and
$\lambda=0.2$, at or near the critical hybridization width
$\Gamma_{c2}\simeq 0.6878956$. Straight lines represent power-law fits.
(a) Static local charge susceptibility $\chi_{c,\loc}(T=0)$ vs
$\Gamma-\Gamma_{c2}$ in the symmetric strong-coupling phase.
(b) Local charge $Q_{\loc}(\phi\to0, T=0)$ vs $\Gamma_{c2}-\Gamma$ in the
local-charge phase. Inset: Continuous vanishing of $Q_{\loc}(\phi\to 0,T=0)$
as $\Gamma$ approaches $\Gamma_{c2}$ from below.
(c) Local charge $Q_{\loc}(\Gamma=\Gamma_{c2}, T=0)$ vs local electric
potential $\phi$.
(d) Static local charge susceptibility
$\chi_{c,\loc}(\Gamma=\Gamma_{c2})$ vs temperature $T$.}
\end{figure}

We have seen above that the NRG spectrum and low-temperature thermodynamics
at the \Cc\ fixed point are related to those at the \Cs\ fixed point by
interchange of spin and charge degrees of freedom. One therefore expects to
be able to probe the critical properties via the response to a local
electric potential $\phi$ that enters the model through an additional
Hamiltonian term
\begin{equation}
\label{phi:def}
\H_{\phi} = \phi(\n_d - 1).
\end{equation}
Comparison with Eq.\ \eqref{H_imp} shows that $\phi$ is equivalent to a
shift in $\dd$ (or $\Ed$).
The order parameter should be the $\phi\to 0$ limiting value of the
local charge
\begin{equation}
Q_{\loc} = \langle \n_d - 1\rangle,
\end{equation}
and the order-parameter susceptibility should be the static local charge
susceptibility
\begin{equation}
\chi_{c,\loc} = -\lim_{\phi\to 0}\frac{Q_{\loc}}{\phi}.
\end{equation}
In the vicinity of the critical point $\Gamma=\Gamma_{c2}$, one expects the
following critical behaviors:
\begin{subequations}
\label{exponents2}
\begin{align}
Q_{\loc}(\Gamma<\Gamma_{c2}; \phi\to0, T=0)
  &\propto (\Gamma_{c2}-\Gamma)^{\beta_2}, \\
\chi_{c,\loc}(\Gamma>\Gamma_{c2}; T=0) &
\propto (\Gamma-\Gamma_{c2})^{-\gamma_2}, \\
Q_{\loc}(\phi; \Gamma=\Gamma_{c2}, T=0) &\propto |\phi|^{1/\delta_2}, \\
\chi_{c,\loc}(T; \Gamma=\Gamma_{c2}) &\propto T^{-x_2}.
\end{align}
\end{subequations}
These expectations are borne out by the NRG results, as illustrated in
Fig.\ \ref{fig:small-r:local_c} for the case $r=0.4$, $\lambda=0.2$ treated
in Fig.\ \ref{fig:small-r:NRGspec_c}.

\subsubsection{Comparison between weak and strong bosonic coupling}

\begin{figure}[t]
\centering
\includegraphics[angle=270,width=0.95\columnwidth]{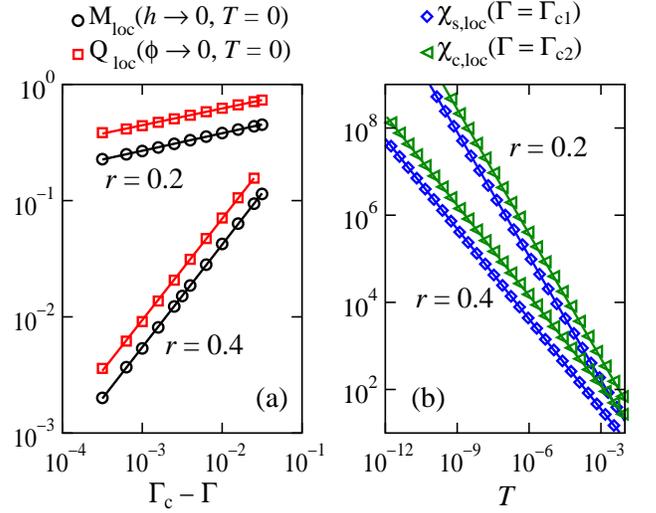}
\caption{\label{fig:small-r:beta+x}
(Color online) Comparison of the static local responses of the
particle-hole-symmetric pseudogap Anderson-Holstein model near its critical
points \Cs\ and \Cc:
(a) Dependence of the local magnetization $M_{\loc}(h\to0, T=0)$ on
$\Gamma_{c1}-\Gamma$ and of the local charge $Q_{\loc}(\phi\to0, T=0)$ on
$\Gamma_{c2}-\Gamma$, for $U=-2\Ed=0.5$, $\omega_0=0.1$, $\lambda=0.05$
(magnetic response) or 0.2 (charge response), and band exponents $r=0.2$ and
$r=0.4$.
(b) Static local spin susceptibility $\chi_{s,\loc}(T;\Gamma=\Gamma_{c1})$
and static local charge susceptibility $\chi_{c,\loc}(T;\Gamma=\Gamma_{c2})$
vs temperature $T$. All parameters other than $\Gamma$ take the same values
as in (a).
Straight lines represent power-law fits.}
\end{figure}

Figure \ref{fig:small-r:beta+x}(a) superimposes the variation with $\Gamma$ of
the order parameter in the vicinity of the \Cs\ and \Cc\ critical points for
two representative band exponents, $r=0.2$ and $0.4$.
The equality of the slopes of the log-log plots at the spin- and charge-sector
QPTs shows that $\beta_1=\beta_2$. Similarly,
Fig.\ \ref{fig:small-r:beta+x}(b) shows that the temperature variation of the
order-parameter susceptibilities is consistent with $x_1=x_2$.
Indeed, for each value of $r$ that we have examined, we find that all critical
exponents at \Cc\ are indistinguishable (within our estimated errors) from the
corresponding exponents at \Cs\ and at the critical point of the pseudogap
Kondo model (as given in Table I of Ref.\ \onlinecite{Ingersent:02}). This
leads us to conclude that all three critical points lie in the same
universality class.

\subsection{Renormalization-group flows}

\begin{figure}[t]
\centering
\includegraphics[width=0.7\columnwidth]{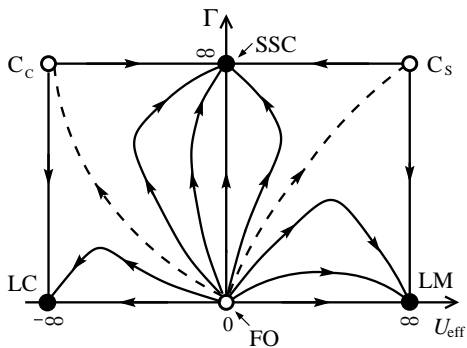}
\caption{\label{fig:small-r:RG_flow}
Schematic renormalization-group flows on the $\Ueff$-$\Gamma$ plane for the
particle-hole-symmetric pseudogap Anderson-Holstein model with a band exponent
$0<r<\half$. Trajectories with arrows represent the flow of the couplings with
decreasing temperature. Dashed lines connecting unstable fixed points (open
circles) separate the basins of attraction of stable fixed points (filled
circles). The right and left dashed lines represent the boundary values
$\Gamma_{c1}$ and $\Gamma_{c2}$ defined in Eqs.\ \eqref{Gamma_c1_vs_lambda}
and \eqref{Gamma_c2_vs_lambda}, respectively. The asymmetry of the flows
about the line $\Ueff=0$ stems from differing symmetries of the model in the
spin and charge sectors. See the text for a discussion of each fixed point.}
\end{figure}

The essential physics of the particle-hole-symmetric pseudogap
Anderson-Holstein model can be summarized in the schematic
renormalization-group flow diagram shown in Fig.\ \ref{fig:small-r:RG_flow},
which applies to all band exponents in the range $0 < r < \half$. Arrows
indicate the evolution of the effective Coulomb interaction $\Ueff$ and the
hybridization width $\Gamma$ with increasing NRG iteration number $N$, i.e.,
under progressive reduction of the temperature $T\simeq D\Lambda^{-N/2}$.  The
high-temperature limit of the model is governed by the \textit{free-orbital}
(FO) fixed point, corresponding to bare model parameters $U=\Gamma=\lambda=0$
and a Fermi-level phase shift $\delta_0(0)=0$.
Dashed lines mark the separatrices between the basins of attraction of the
local-moment (LM), local-charge (LC), and symmetric strong-coupling (SSC) fixed
points described above. Flow along each separatrix is from the free-orbital
fixed point towards one or other of two quantum critical points---either the
conventional spin-sector critical point \Cs\ reached for $\Ueff > 0$, or
its charge analog \Cc\ reached for $\Ueff < 0$.

A renormalization-group fixed-point structure equivalent to that described in
the preceding paragraph has been presented previously\cite{Fritz:04} for the
pseudogap Anderson model under the assumption that the bare on-site Coulomb
interaction $U$ may be taken to be positive or negative. Indeed, many of the
universal properties of the pseudogap Anderson-Holstein model presented in this
section---particularly ones associated with the quantum critical points \Cs\
and \Cc---reproduce those of this extended pseudogap Anderson model.

However, we emphasize that the particle-hole-symmetric Anderson and
Anderson-Holstein models have different symmetries and are therefore not
trivially related to one another. The pseudogap Anderson Hamiltonian exhibits
exact SU(2) spin and isospin (charge) symmetries, and all physical properties at
a point $(U,\Gamma)$ in the diagram analogous to Fig.\ \ref{fig:small-r:RG_flow}
[see Fig.\ 1(b) of Ref.\ \onlinecite{Fritz:04}] map exactly to the properties
at $(-U,\Gamma)$ under the interchange of spin and charge degrees of freedom.
No such mapping holds in the pseudogap Anderson-Holstein model, where the
Hamiltonian has full SU(2) spin symmetry but only a discrete charge symmetry.
This distinction leads, for instance, to the critical hybridization width
$\Gamma_{c1}$ having a sublinear dependence on $\Ueff$
[Eq.\ \eqref{Gamma_c1_vs_lambda}] whereas its counterpart
$\Gamma_{c2}$ is superlinear in $\Ueff$ [Eq.\ \eqref{Gamma_c2_vs_lambda}].
The equivalence of the critical points \Cs\ and \Cc\ under spin-charge
interchange signals the emergence of a higher SU(2) isospin symmetry at
both these renormalization-group fixed points.

For $r\ge\half$, we find that (just as in the pseudogap Anderson
model\cite{Fritz:04}) the symmetric strong-coupling fixed point of the
pseudogap Anderson-Holstein model is unstable with respect to any breaking of
degeneracy between the four impurity levels, i.e., to any $\Ueff\ne 0$. As a
result, the $\Ueff$-$\Gamma$ plane is divided into just two phases:
local-moment for all $\Ueff>0$ and local-charge for all $\Ueff<0$.

\section{Results: General Model with Band Exponent $0<r<1$}
\label{sec:small-r}

This section treats the pseudogap Anderson-Holstein model with a band
exponent $0<r<1$ when either (the discrete) particle-hole symmetry is broken
by a value $\dd \equiv \Ed + \half U \ne 0$ or (the continuous) spin-rotation
invariance is removed by a nonvanishing local magnetic field $h$ [defined in
Eq.\ \eqref{h:def}].
It is found that increasing $|\dd|$ or $|h|$ can drive the system from the
local-moment or local-charge phase into one of several strong-coupling phases
that are not present in the baseline case $\dd = h = 0$. The transitions
between these phases take place at interacting quantum critical points in the
same universality class as the asymmetric critical points of the pseudogap
Anderson model.

All numerical results presented in this section were obtained for an impurity
with $U = 0.5$, for a bosonic energy $\omega_0 = 0.1$, and for NRG
discretization parameter $\Lambda = 3$.

\subsection{Phase boundaries}

\subsubsection{Weak bosonic coupling}

Figure~\ref{fig:phase_bound_spin} plots phase boundaries of the pseudogap
Anderson-Holstein model on the $\Gamma$-$\dd$ plane for zero magnetic
field, for band exponents $r=0.4$ (left) and $0.6$ (right), and for three
bosonic couplings $\lambda = 0$, 0.1, and $0.1414$ that can all be associated
[via Eq.\ \eqref{Ubar:def}] with effective Coulomb interactions $\Ubar>0$.
A value $\dd\ne 0$ breaks particle-hole symmetry but leaves in place the
SU(2) spin symmetry. The system remains in the local-moment phase for
$|\dd|<\frac{1}{2}\Ubar$ and $\Gamma < \Gamma_{c1}(r,U,\dd,\lambda) \equiv
\Gamma_{c1}(r,U,-\dd,\lambda)$. Otherwise it lies in one of the
strong-coupling phases described in Sec.\ \ref{subsubsec:PAM}: symmetric
strong-coupling for $\dd=0$, \ASCm\ for $\dd > 0$, or \ASCp\ for
$\dd < 0$. Just as in the pseudogap Anderson model,\cite{Gonzalez-Buxton:98}
the symmetric strong-coupling phase can be reached only for $r<\half$; for
$r\ge\half$, the symmetric strong-coupling fixed point is unstable and
$\Gamma_{c1}(r,U,\dd)$ diverges as $\dd$ approaches zero. The
contraction of the local-moment phase (i.e., the reduction of $\Gamma_{c1}$)
with increasing $\lambda$ and/or $|\dd|$ can be attributed to one or both
of the $n_d\ne 1$ impurity levels being drawn down in energy closer to the
$n_d=1$ ground states. This energy shift enhances indirect spin-flip scattering
between the ground states via the excited states and favors conduction-band
quenching of the impurity degrees of freedom.

\begin{figure}[t]
\centering
\includegraphics[angle=270,width=0.95\columnwidth]{Fig11.eps}
\caption{\label{fig:phase_bound_spin}
(Color online) Phase boundaries of the zero-field pseudogap Anderson-Holstein
model on the $\Gamma$-$\dd$ plane for weak bosonic couplings
$\lambda<\lambda_0=0.15812(1)$ and band exponent $r=0.4$ (left), $r=0.6$
(right). Data calculated for $U=0.5$, $h=0$, $\omega_0=0.1$, and the three
values of $\lambda$ listed in the legend.}
\end{figure}

\subsubsection{Strong bosonic coupling}

\begin{figure}[t]
\centering
\includegraphics[angle=270,width=0.95\columnwidth]{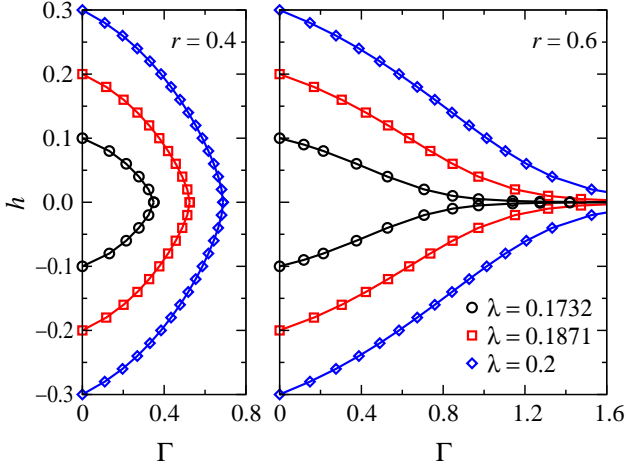}
\caption{\label{fig:phase_bound_charge}
(Color online) Phase boundaries of the particle-hole-symmetric pseudogap
Anderson-Holstein model on the $\Gamma$-$h$ plane for strong bosonic couplings
$\lambda>\lambda_0=0.15812(1)$ and band exponent $r=0.4$ (left), $r=0.6$
(right). Data calculated for $U=-2\epsilon_d=0.5$, $\omega_0=0.1$, and the
three values of $\lambda$ listed in the legend.}
\end{figure}

Figure~\ref{fig:phase_bound_charge} plots phase boundaries of the
particle-hole-symmetric pseudogap Anderson-Holstein model on the $\Gamma$-$h$
plane for $r=0.4$ (left) and $0.6$ (right), and for three strong bosonic
couplings in the range $\lambda>\lambda_0$. Here, Eq.\ \eqref{Ubar:def} gives
$\Ubar<0$, and for local magnetic field $h=0$ the lowest-energy impurity states
are spinless but have a charge $Q=\pm 1$. Application of a field $h\ne 0$
destroys SU(2) spin symmetry but preserves particle-hole symmetry. The field
pulls one or other of the $Q=0$ excited states down in energy, thereby enhancing
virtual scattering between the $Q=\pm 1$ states. The model remains in the
local-charge phase for $|h|<|\Ubar|$ and
$\Gamma<\Gamma_{c2}(r,U,h,\lambda)\equiv\Gamma_{c2}(r,U,-h,\lambda)$. Otherwise
it lies in the symmetric strong-coupling phase (for $r<\half$ and $h=0$) or in
one of two new asymmetric strong-coupling phases: \ASCd\ with ground-state spin
$z$ component $S_z = -\half$ for $h > 0$, or \ASCu\ with $S_z = \half$ for
$h < 0$. For $r\ge\half$, $\Gamma_{c2}(r,U,h,\lambda)$ diverges as $h$
approaches zero, a consequence of the instability of the symmetric
strong-coupling fixed point in this range of band exponents.

Although the two panels of Fig.\ \ref{fig:phase_bound_charge} look quite
similar to their counterparts in Fig.\ \ref{fig:phase_bound_spin}, there are
small departures such as the absence from $\Gamma_{c1}$ versus $\dd$ for $r=0.6$
of points of inflection corresponding to those seen in $\Gamma_{2c}$ versus $h$.
These differences presumably arise from the differing symmetries of the two
cases.

\subsection{Impurity thermodynamic properties}

\begin{figure}[t]
\centering
\includegraphics[angle=270,width=0.95\columnwidth]{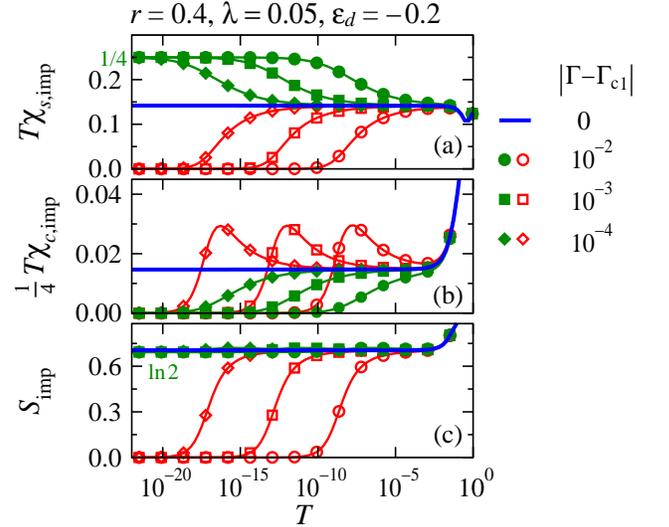}
\caption{\label{fig:r=0.4_asymm:thermo_s}
(Color online) Thermodynamic properties of the particle-hole-asymmetric
pseudogap Anderson-Holstein model in zero magnetic field near the spin-sector
critical point \Cm: Temperature dependence of the impurity contribution to
(a) the static spin susceptibility $\chi_{s,\imp}$ multiplied by temperature,
(b) the static charge susceptibility $\chi_{c,\imp}$ multiplied by temperature,
and (c) the entropy $S_{\imp}$,
for $r=0.4$, $\Ed=-0.2$, $U=0.5$, $\omega_0 = 0.1$,
$\lambda=0.05<\lambda_0\simeq 0.158$, and the seven values of
$\Gamma-\Gamma_{c1}$ labeled in the legend. Filled (open) symbols connected by
guiding lines represent data in the local-moment (\ASCm) phase, while thick lines
without symbols show the critical properties at \Cm. $N_s = 3\,000$ states
were retained after each NRG iteration.}
\end{figure}

\subsubsection{Weak bosonic coupling}

Figure \ref{fig:r=0.4_asymm:thermo_s} plots the temperature dependence of
$T\chi_{s,\imp}$, $\frac{1}{4}T\chi_{c,\imp}$, and $S_{\imp}$ for $r=0.4$,
$U=0.5$, $\Ed = -0.2$ (or $\dd = 0.05 > 0$), $\lambda=0.05$, and seven
values of $\Gamma$ straddling $\Gamma_{c1}$. The low-temperature limiting
behaviors in the local-moment phase ($\Gamma<\Gamma_{c1}$) are identical to
those found at particle-hole symmetry (see Fig.\ \ref{fig:r=0.4_symm:thermo_s}).
In the \ASCm\ phase ($\Gamma>\Gamma_{c1}$), however, the $T=0$ properties
$T\chi_{s,\imp}=\frac{1}{4}T\chi_{c,\imp}=S_{\imp}=0$ indicate complete
quenching of the impurity degrees of freedom, in contrast to the partial
quenching found in the symmetric strong-coupling phase. Exactly at $\Gamma =
\Gamma_{c1}$, the low-temperature properties $T\chi_{s,\imp}\simeq 0.1419$,
$\frac{1}{4}T\chi_{c,\imp}\simeq 0.0147$, and $S_{\imp}\simeq 0.705$ can be
taken to characterize the critical point \Cm\ separating the two stable phases.
These properties coincide with those found for $\dd < 0$ at the critical
point \Cp\ between the local-moment and \ASCp\ phases, and also with those of
the asymmetric critical points of the pseudogap Kondo or Anderson
models;\cite{Gonzalez-Buxton:98} however, for $r>r^*\simeq 3/8$ these properties
differ from from those of the corresponding symmetric critical point \Cs. (Our
observation that the asymmetric critical value of $S_{\imp}$ is slightly greater
than $\ln 2$ is consistent with Refs.\ \onlinecite{Kircan:04} and
\onlinecite{Fritz:04}.)

For $\Gamma$ close to $\Gamma_{c1}$, the crossover of the thermodynamic
properties away from their critical values can be used to define a crossover
scale $T^*_1$ that obeys Eq.\ \eqref{nu1}. Table \ref{tab:small-r_asymm} lists
the correlation-length exponent $\nu_1(r)$ obtained for five values of $r$.

\begin{figure}[t]
\centering
\includegraphics[angle=270,width=0.95\columnwidth]{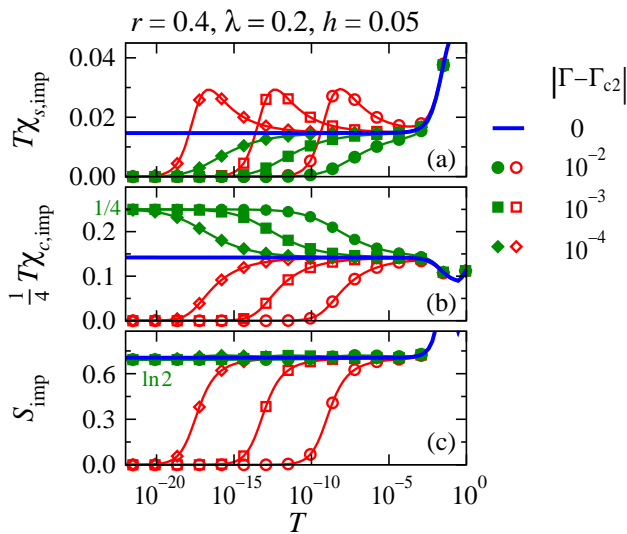}
\caption{\label{fig:r=0.4_symm_h:thermo_c}
(Color online) Thermodynamic properties of the particle-hole-symmetric
pseudogap Anderson-Holstein model in nonzero local magnetic field near the
charge-sector critical point \Cd:
Temperature dependence of the impurity contribution to
(a) the static spin susceptibility $\chi_{s,\imp}$ multiplied by temperature,
(b) the static charge susceptibility $T\chi_{c,\imp}$ multiplied by temperature,
and (c) the entropy $S_{\imp}$, for $r=0.4$, $U=-2\Ed=0.5$, $h=0.05$,
$\omega_0 = 0.1$, $\lambda=0.2>\lambda_0\simeq 0.158$, and the seven values
of $\Gamma-\Gamma_{c2}$ labeled in the legend. Filled (open) symbols connected
by guiding lines represent data in the local-charge  (\ASCd) phase, while thick
lines without symbols show the critical properties at \Cd. $N_s = 3\,000$ states
were retained after each NRG iteration.}
\end{figure}

\subsubsection{Strong bosonic coupling}

Figure \ref{fig:r=0.4_symm_h:thermo_c} plots the temperature dependence of
$T\chi_{s,\imp}$, $\frac{1}{4}T\chi_{c,\imp}$, and $S_{\imp}$ for $r=0.4$,
$U=-2\Ed=0.5$ (i.e., $\dd = 0$), $h=0.05$, $\lambda=0.2$, and
various $\Gamma$ straddling the critical value $\Gamma_{c2}$. Just as was
found at particle-hole symmetry, the thermodynamic properties for
$\lambda>\lambda_0$ are related to those for $\lambda<\lambda_0$ by
interchange of spin and charge degrees of freedom. For the cases shown
in Fig.\ \ref{fig:r=0.4_symm_h:thermo_c}, it is the charge susceptibility
that most clearly distinguishes the critical point \Cd\
($\frac{1}{4}T\chi_{c,\imp}\simeq 0.1419$) from the local-charge and \ASCd\
phases  ($\frac{1}{4}T\chi_{c,\imp}=\frac{1}{4}$ and $0$, respectively).

\subsection{Local response and universality class}

In the vicinity of the quantum critical points \Cpm\ separating the local-moment
and \ASCpm\ phases, the local spin responses exhibit power-law behaviors
described by Eqs.\ \eqref{nu1} and \eqref{exponents1}. Table
\ref{tab:small-r_asymm} lists critical exponents at \Cpm\ obtained for five
different values of $r$. To within their estimated accuracy, the exponents
obey the hyperscaling relations Eq.\ \eqref{hyper}, providing evidence for the
interacting character of the critical points.

Comparison between Tables~\ref{tab:small-r} and \ref{tab:small-r_asymm} shows
that the symmetric critical point \Cs\ and its asymmetric counterparts \Cpm\
have the same low-temperature physics for $r=0.2$ and $0.3$, but not for
$r=0.4$. This pattern is consistent with the pseudogap Kondo
model,\cite{Gonzalez-Buxton:98} where the \Cs\ and \Cpm\ critical points are
identical for $0<r<r^*\simeq 3/8$ but distinct for $r^*\lesssim r<1$.
In the latter range, the exponents listed in Table~\ref{tab:small-r_asymm}
coincide to within small errors with those for the particle-hole-asymmetric
pseudogap Kondo model given in Table II of Ref.\ \onlinecite{Ingersent:02}.

\begin{table}[t]
\caption{\label{tab:small-r_asymm}
Exponents describing the local spin response at the critical points \Cpm\ of
the particle-hole-asymmetric pseudogap Anderson-Holstein model, evaluated
for five band exponents $r$. The critical exponents are defined in Eqs.\
\protect\eqref{nu1} and \protect\eqref{exponents1}. A number in parentheses
indicates the estimated random error in the last digit of each exponent.}
\begin{tabular*}{.98\columnwidth}{@{\extracolsep{\fill}}llllll}
\hline\hline \\[-2ex]
\multicolumn{1}{c}{$r$} & \multicolumn{1}{c}{$\nu_1$} &
\multicolumn{1}{c}{$\beta_1$} & \multicolumn{1}{c}{$1/\delta_1$} &
\multicolumn{1}{c}{$x_1$} & \multicolumn{1}{c}{$\gamma_1$} \\[.2ex]
\hline \\[-2ex]
0.2 &   6.22(1)  & 0.15(1)  & 0.02630(2) & 0.9488(2) & 5.85(6) \\
0.3 &   5.14(1)  & 0.34(1)  & 0.07364(1) & 0.8629(3) & 4.41(3) \\
0.4 &   4.29(1)  & 0.59(1)  & 0.1569(1)  & 0.7275(3) & 3.12(2) \\
0.6 & 1.78(1)  & 0.188(1) & 0.1173(2)  & 0.7896(4) & 1.41(1) \\
0.8 & 1.27(1)  & 0.079(1) & 0.0644(5)  & 0.879(1)  & 1.10(1) \\[.2ex]
\hline\hline
\end{tabular*}
\end{table}

In the vicinity of the critical points \Cud\ marking the transitions from the
local-charge phase to the \ASCud\ phases, the local charge responses exhibit
power-law behaviors described by critical exponents equal (within small errors)
to the local-spin exponents of the \Cpm\ critical points of the pseudogap
Kondo and pseudogap Anderson-Holstein models. We are led to conclude that
these critical points all lie in the same universality class.
Given the different symmetries of the Anderson-Holstein model under spin and
isospin rotation, the equivalence of the \Cpm\ and \Cud\ critical points under
spin-charge interchange is a nontrivial finding, distinct from the equivalence
of the \Cs\ and \Cc\ critical points at particle-hole symmetry.

\section{Results: Double Quantum Dots With $U_2 = 0$}
\label{sec:r=2}

In the pseudogap Kondo and Anderson models, the interacting quantum critical
points found for band exponents $0<r<1$ are replaced for $r\ge 1$ by
first-order QPTs that arise from renormalized level
crossings between spin-doublet and spin-singlet ground states of the
impurity.\cite{Fritz:04} Similar behavior is expected in the pseudogap
Anderson-Holstein model. This section focuses on the particular case $r=2$
that is of particular interest because it has a possible realization in
double quantum dots. Below we present results not only for the impurity
contributions to thermodynamic properties but also for the linear
conductance of such a double-dot system in the vicinity of its spin- and
charge-sector QPTs.

\subsection{Pseudogapped effective model for double quantum dots}

The motivation for focusing on the case $r=2$ comes from theoretical
studies\cite{Dias:06,Dias:08,Dias:09} of two lateral quantum dots coupled in
parallel to left ($L$) and right ($R$) leads, and gated in such a manner that
the low-energy physics is dominated by just one single-particle state on each
dot. It is assumed that one of the dots (dot 1) is small and hence strongly
interacting, while the other (dot 2) is larger, has a negligible charging
energy, and can be approximated as a noninteracting resonant level.
This setup can be described by the two-impurity Anderson Hamiltonian
\begin{align}
\label{H_DD}
\H_{DD}
&= \sum_{j,\s}\veps_j \, \n_{j\s} + U_1\n_{1\up}\n_{1\dn}
   + \sum_{\ell,\bk,\s}\veps_{\ell\bk}^{\pdag}c^{\dag}_{\ell\bk\s}
   c^{\pdag}_{\ell\bk\s} \nonumber \\
& + \sum_{j,\ell,\bk,\s} V_{j\ell} \bigl( d^{\dag}_{j\s}
  c^{\pdag}_{\ell\bk\s} + \text{H.c.} \bigr).
\end{align}
Here, $d_{j\s}$ annihilates an electron of spin $z$ component $\s$ and
energy $\veps_j$ in the dot $j$ ($j=1$, 2), $\n_{j\s} =
d_{j\s}^{\dag} d_{j\s}$ is the number operator for such
electrons, and $c_{\ell\bk\s}$ annihilates an electron of spin $z$
component $\s$ and energy $\veps_{\ell\bk}$ in lead $\ell$
($\ell = L$, $R$). For simplicity, the leads are assumed to have the same
dispersion $\veps_{\ell\bk}=\Ek$ corresponding to a ``top-hat''
density of states $\rho(\veps)=\rho_0\Theta(D-|\veps|)$ with
$\rho_0=(2D)^{-1}$, and to hybridize symmetrically with the dots so that
$V_{jL}=V_{jR}$. Under these conditions, the dots couple only to the
symmetric combination of lead electrons annihilated by
$c_{\bk\s}=(c_{L\bk\s}+c_{R\bk\s})/\sqrt{2}$ with effective
hybridization matrix elements $V_j=\sqrt{2}V_{j\ell}$.

A key feature of Eq.\ \eqref{H_DD} is the vanishing of the dot-2 Coulomb
interaction $U_2$ associated with a Hamiltonian term $U_2 \n_{2\up}
\n_{2\dn}$. This allows one to integrate out dot 2 to yield an effective
Anderson model for a single impurity characterized by a level energy $\veps_1$,
an on-site interaction $U_1$, and a hybridization function \cite{Dias:06}
\begin{equation}
\label{Gamma_dot}
\Gam_1(\veps) =
  \frac{(\veps-\veps_2)^2}{(\veps-\veps_2)^2+\Gamma_2^2} \, \Gamma_1 \,
  \Theta(D-|\veps|) ,
\end{equation}
where $\Gamma_j=\pi\rho_0 V_j^2$ for $j=1$, 2. The presence of dot 2 in the
original model manifests itself here as a Lorentzian hole in
$\Gam_1(\veps)$ of width $\Gamma_2$ centered on $\veps=\veps_2$.
For $\veps_2=0$ (a condition that might be achieved in practice by tuning
a plunger gate voltage on dot 2), $\Gam_1(\veps)\propto\veps^2$
in the vicinity of the Fermi energy, providing a realization of the $r=2$
pseudogap Anderson model.\cite{Dias:06}

In the remainder of this section, we consider the double-dot device
introduced in Ref.\ \onlinecite{Dias:06}, augmented by a Holstein coupling
between dot 1 and local bosons. Such a system, modeled by a Hamiltonian
$\H_{DD}+\omega_0 a^{\dag} a^{\pdag} + \lambda (\n_1-1)(a^{\pdag}+a^{\dag})$,
can be mapped (following Ref.\ \onlinecite{Dias:06}) onto the effective
single-impurity model
\begin{align}
\label{H_DD:PAH}
\H
&= \sum_{\s} \veps_1 \n_1 + U_1 \n_{\up} \n_{\dn} +
   \sum_{\bk,\s}\veps^{\pdag}_{\bk}c^{\dag}_{\bk\s}
   c^{\pdag}_{\bk\s} + \omega_0 a^{\dag} a^{\pdag} \nonumber\\
&+ \sum_{\bk,\s} V_1 \bigl( d^{\dag}_{\s}c^{\pdag}_{\bk\s}+
   \text{H.c.} \bigr) + \lambda (\n_1 - 1) \bigl(a^{\pdag}\! + a^{\dag} \bigr)
\end{align}
with the hybridization function $\Gam_1(\veps) = \pi N_k^{-1} \sum_{\bk} V_1^2
\delta(\veps-\Ek)$ as defined in Eq.\ \eqref{Gamma_dot}.

All numerical results presented in the remainder of this section were obtained
using the effective one-impurity pseudogap Anderson-Holstein model
[Eq.\ \eqref{H_DD:PAH}] for a strongly interacting dot 1 having $U_1=0.5$ and
for a bosonic frequency $\omega_0=0.1$. The NRG calculations were performed for
a discretization parameter $\Lambda=2.5$.

\subsection{Phase boundaries}
\label{subsec:r=2:phase_diagram}

\subsubsection{Weak bosonic coupling}

\begin{figure}[t]
\centering
\includegraphics[angle=270,width=0.65\columnwidth]{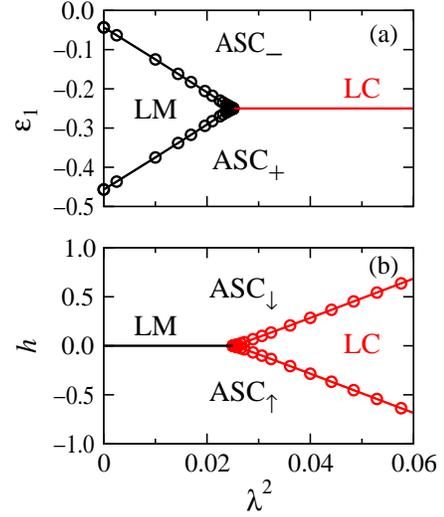}
\caption{\label{fig:r=2:phase_diag}
(Color online) Phase diagrams of a $U_2=0$ double-quantum-dot device with
$U_1=0.5$, $\Gamma_1=0.05$, $\veps_2=0$, $\Gamma_2=0.02$, and $\omega_0=0.1$:
(a) The $\lambda^2$-$\veps_1$ plane for $h=0$.
(b) The $\lambda^2$-$h$ plane for $\veps_1=-\half U_1$.
Note the near-linearity of the phase boundaries in each panel when plotted
against the square of the bosonic coupling.}
\end{figure}

Figure \ref{fig:r=2:phase_diag}(a) shows the phase diagram of the $U_2=0$
double-quantum-dot device, as mapped to the pseudogap Anderson-Holstein model,
on the $\lambda^2$-$\veps_1$ plane in the absence of any magnetic field. For
$\lambda<\lambda_0\simeq0.1582848$, decreasing the dot-$1$ energy starting from
a large positive value drives the system from the \ASCm\ phase to the
local-moment phase (LM) at $\veps_1=\veps_{1,c}^+ \le 0$, and then from the
local-moment phase to the \ASCp\ phase at
$\veps_1=\veps_{1,c}^- = -U_1 - \veps_{1,c}^+$. For $\lambda>\lambda_0$,
the system instead lies in the \ASCm\ phase for all $\veps_1 > -\half U_1$, in
the \ASCp\ phase for all $\veps_1 < -\half U_1$, and in the local-charge phase
(LC) only along the line $\veps_1=-U_1/2$ of strict particle-hole symmetry.

One of the most notable features of Fig.\ \ref{fig:r=2:phase_diag}(a) is the
linear dependence of $\veps_{1,c}^{\pm}$ on $\lambda^2$, which implies a linear
dependence on the polaron energy $\Ep=\lambda^2/\omega_0$. Since $\veps_{1,c}^-
= -U_1-\veps_{1,c}^+$, it suffices to focus on the phase boundary between the
\ASCm\ and LM phases. In the atomic limit $\Gamma=0$, one expects this boundary
to be defined by the degeneracy of the $n_d=0$ and $n_d=1$ impurity levels,
i.e., the point where the renormalized dot-1 level energy [cf.\ Eq.\
\eqref{Edbar:def}] satisfies $\bar{\veps}_1 \equiv \veps_1+\Ep = 0$, a
condition that implies $\veps_{1,c}^+=-\Ep$.

The location of the phase boundary for $\Gamma>0$ can be estimated using the
poor-man's scaling equations discussed in Sec.\ \ref{subsec:scaling}. Equation
\eqref{tildeGamma} implies that for $r>1$, the effective value of the
dimensionless scattering width $\tG/\tD = (\tD/D)^{r-1} (\Gamma/D)$ decreases
monotonically under reduction of the half-bandwidth from $D$ to $\tD$. If
$\Gamma<D$, this decrease rules out the possibility of entry into the
mixed-valence regime under the criteria laid out at the end of Sec.\
\ref{subsec:scaling}. Moreover, the decrease of $\tG$ is so rapid that any entry
to the local-moment regime and subsequent mapping to the pseudogap Kondo problem
will yield a sub-critical exchange coupling, placing the system in the local-moment
phase. Under these circumstances, the boundary between the local-moment phase and
the asymmetric strong-coupling phase \ASCm\ [see Sec.\ \ref{subsubsec:PAM} and in
particular Fig.\ \ref{fig:PAM_phase}(b)] is effectively determined by the condition
$\tEd(\tD=0)=0$ for a level crossing between the renormalized energies of the empty
and singly occupied impurity configurations. With the approximation $U=\infty$, and
using the expansion
\begin{equation}
S(a,x) \simeq 1 - \frac{x}{a+1} + O(x^2) \quad \text{for } x>0,
\end{equation}
Eq.\ \eqref{Ed:scaling} can be integrated to yield
\begin{equation}
\label{Edc}
\veps_{d,c}^+ \simeq -\frac{\Gamma}{2\pi}-\alpha_1\,\Ep ,
\end{equation}
with
\begin{equation}
\alpha_1 = 1 - \biggl( 1-\frac{\omega_0}{D}\,\ln\frac{D+\omega_0}{\omega_0}\biggr)
  \frac{\Gamma}{\pi D}
\end{equation}
The predicted value $\alpha_1=0.986$ is in good agreement with the one
$\alpha_1=0.988$ that describes NRG results for the Anderson-Holstein model
with $U=0.5$, $\Gamma=0.05$, and a pure $r=2$ power-law hybridization function
(data not shown). The phase boundary for the mapped double-quantum-dot system
plotted in Fig.\ \ref{fig:r=2:phase_diag}(a) can be fitted with a reduced value
$\alpha_1=0.80$ that can be attributed to the fact that the hybridization
function $\Gam_1(\veps)$ in Eq.\ \eqref{Gamma_dot} assumes a power-law form
only for $|\veps|\ll\Delta_2\ll D$.

\subsubsection{Strong bosonic coupling}

Figure \ref{fig:r=2:phase_diag}(b) shows the phase diagram of the $U_2=0$
double-quantum-dot system on the $\lambda^2$-$h$ plane at particle-hole
symmetry. For $\lambda>\lambda_0$, decreasing the local magnetic field from a
large positive value takes the system from the \ASCd\ phase to the local-charge
phase (LC) at $h=h_c > 0$, and then from the local-charge phase to the \ASCu\
phase at $h = -h_c$. For $\lambda<\lambda_0$, by contrast, the system is in the
\ASCd\ or \ASCu\ phase for $h>0$ or $h<0$, respectively, and in the local-charge
phase only along the line $h=0$.

The phase boundaries shown in Fig.\ \ref{fig:r=2:phase_diag}(b) are nearly
linear in $\lambda^2-\lambda_0^2$ or, equivalently, linear in $2\Ep-U$.
In the atomic limit $\Gamma=0$, one expects the boundary between the \ASCd\ and
LC phases to be located at the point where the singly occupied $S_z=-\half$
impurity state crosses energies with the degenerate pair of impurity states
having $n_d=0$, $2$, i.e., to satisfy $-\half h_c=\half\Ubar$ or $h_c=2\Ep-U$.
For $\Gamma>0$, a generalization of the poor-man's scaling analysis of Sec.\
\ref{subsec:scaling} to incorporate the local field would be expected to yield
corrections to this result along lines similar to the corrections found in the
regime $\lambda<\lambda_0$. Empirically, we find that the data in Fig.\
\ref{fig:r=2:phase_diag}(b) for $0<\lambda-\lambda_0\ll\lambda_0$ can be fitted
to the form
\begin{equation}
\label{h_c}
h_c \simeq \alpha_2 (2\Ep - U)
\end{equation}
with $\alpha_2=0.69$. However, a larger value $\alpha_2\simeq 1$ is required to
describe the data points for $\lambda\simeq 2\lambda_0$, indicating that the
critical field is not strictly linear in $2\Ep-U$.

\subsection{Crossover scales}
\label{subsec:r=2_symm:Tstar}

\begin{figure}[t]
\centering
\includegraphics[angle=270,width=0.8\columnwidth]{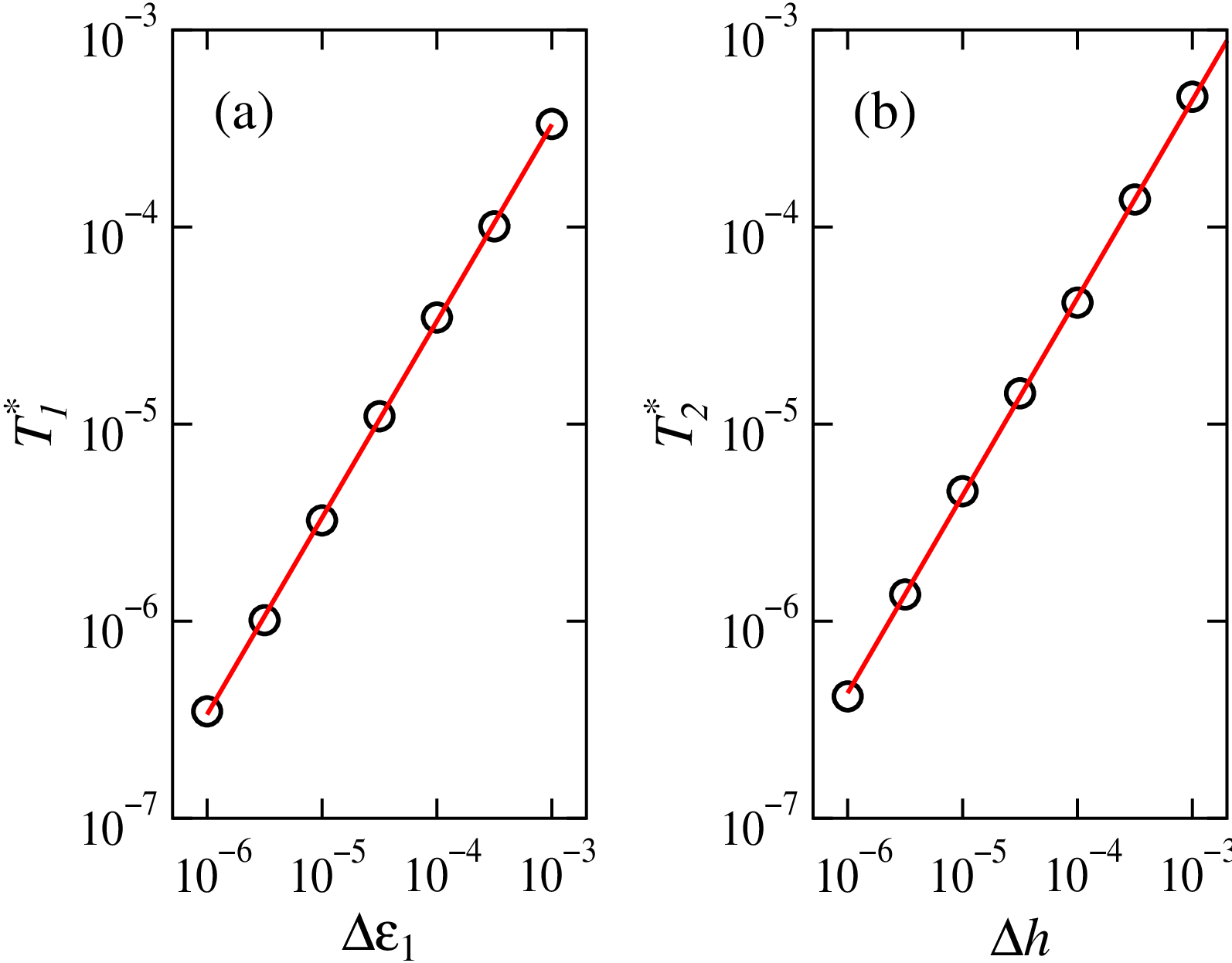}
\caption{\label{fig:r=2:crossover}
(Color online) Crossover temperature scales near QPTs in a $U_2=0$
double-quantum-dot device with $U_1=0.5$, $\Gamma_1=0.05$, $\veps_2=0$,
$\Gamma_2=0.02$, and $\omega_0=0.1$:
(a) $T^*_1$ vs $\Delta\veps_1=\veps_1-\veps^{+}_{1,c}$ in the \ASCm\ phase for
$h=0$ and $\lambda=0.1$.
(b) $T^*_2$ vs $\Delta h=h-h_c$ in the \ASCd\ phase for $\veps_1=-0.25$ and
$\lambda=0.2$.
Straight lines represent power-law fits, which yield the correlation-length
exponents $\nu_1=\nu_2=1$. $N_s=1\,000$ states were retained after each NRG
iteration.}
\end{figure}

As discussed in Sec.\ \ref{subsec:small-r_symm:crossovers}, we can use the NRG
spectrum to identify temperature scales characterizing crossovers between
different renormalization-group fixed points. Figure \ref{fig:r=2:crossover}(a)
plots the crossover scale $T^*_1$ versus $\Delta\veps_1=\veps_1-\veps_{1,c}^+$
on approach to the local-moment phase boundary from the \ASCm\ phase for a weak bosonic
coupling $\lambda=0.1$, while Fig.\ \ref{fig:r=2:crossover}(b) shows the scale
$T^*_2$ versus $\Delta h = h - h_c$ in the \ASCd\ phase near the local-charge phase boundary
for $\lambda = 0.2$. $T^*_1$ and $T^*_2$ vanish at the phase boundaries in the
manner
\begin{equation}
\label{nu1_r=2}
T^*_1 \propto |\veps-\veps^+_{c1}|^{\nu_1}
  \quad \text{as} \quad \veps \to \veps_{1,c}^+,
\end{equation}
and
\begin{equation}
\label{nu2_r=2}
T^*_2 \propto |h-h_c|^{\nu_2} \quad \text{as} \quad h \to h_c,
\end{equation}
with correlation-length exponents $\nu_1=\nu_2=1$. This linear vanishing
of crossover scales is consistent with the level-crossing nature of the QPTs
of the pseudogap Anderson-Holstein model for $r=2$.

\subsection{Impurity thermodynamic properties}
\label{subsec:therm}

As in the cases $r<1$ considered above, the temperature variation of the
impurity contributions\cite{impurity-props} to the static spin and charge
susceptibilities and to the entropy can be used to distinguish the
strong-coupling phases of the $U_2 = 0$ double-quantum-dot system from the
phases with residual local spin or charge degrees of freedom.

\subsubsection{Weak bosonic coupling}

\begin{figure}[t]
\centering
\includegraphics[angle=270,width=0.95\columnwidth]{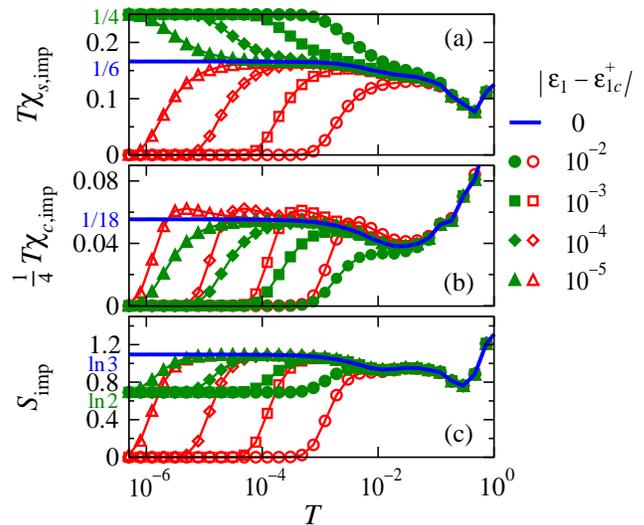}
\caption{\label{fig:r=2:thermo_s}
(Color online) Thermodynamic properties of a $U_2=0$ double-quantum-dot device
near a spin-sector QPT:
Temperature dependence of the impurity contribution to
(a) the static spin susceptibility $\chi_{s,\imp}$ multiplied by temperature,
(b) the static charge susceptibility $\chi_{c,\imp}$ multiplied by temperature,
and (c) the entropy $S_{\imp}$, for $U_1=0.5$,
$\Gamma_1=0.05$, $\veps_2=0$, $\Gamma_2=0.02$, $h=0$, $\omega_0=0.1$,
$\lambda=0.1$, and nine values of $|\veps_1-\veps_{1,c}^+|$ where
$\veps_{1,c}^+\simeq -0.124985$. Properties at the transition (thick lines
without symbols) are those expected at a level crossing between the
local-moment phase (filled symbols) and the \ASCm\ phase. $N_s = 3\,000$
states were retained after each NRG iteration.}
\end{figure}

Figure \ref{fig:r=2:thermo_s} plots the temperature dependence of the impurity
thermodynamic properties $T\chi_{s,\imp}$, $\frac{1}{4}T\chi_{c,\imp}$, and
$S_{\imp}$ for $h=0$, a weak bosonic coupling $\lambda=0.1$, and various values
of $\veps_1$ straddling the upper transition. For
$\veps_1=\veps_{1,c}^+\simeq -0.124985$ (lines without symbols in Fig.\
\ref{fig:r=2:thermo_s}), the low-temperature limiting values
$T\chi_{s,\imp}=1/6$, $\frac{1}{4}T\chi_{c,\imp}=1/18$, and $S_{\imp}=\ln 3$
are those characteristic of the \textit{valence-fluctuation} fixed point: the
point of degeneracy between impurity occupancies $n_d=0$ and $n_d=1$
corresponding to Eq.\ \eqref{H_DD:PAH} with effective couplings
$\veps_1=\Gamma_1=\lambda=0$ and $U_1=\infty$.
If $\veps_1$ deviates slightly from its critical value, the
properties trace their critical behaviors at high temperatures, but cross over
below a scale $T^*_1$ to those either of the local-moment phase, where there
is a residual spin-$\half$ degree of freedom ($T\chi_{s,\imp}=1/4$,
$T\chi_{c,\imp}=0$, and $S_{\imp}=\ln 2$) or of the \ASCm\ phase
($T\chi_{s,\imp}=T\chi_{c,\imp}=S_{\imp}=0$).

\begin{figure}[t]
\centering
\includegraphics[angle=270,width=0.95\columnwidth]{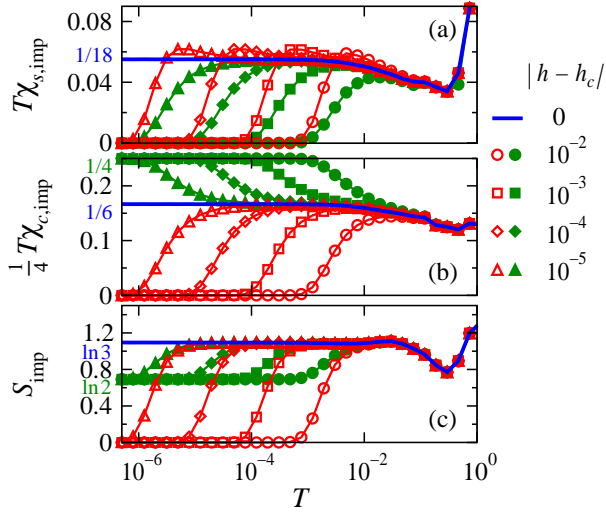}
\caption{\label{fig:r=2:thermo_c}
(Color online) Thermodynamic properties of a $U_2=0$ double-quantum-dot device
near a charge-sector QPT:
Temperature dependence of the impurity contribution to
(a) the static spin susceptibility $\chi_{s,\imp}$ multiplied by temperature,
(b) the static charge susceptibility $\chi_{c,\imp}$ multiplied by temperature,
and (c) the entropy $S_{\imp}$, for $U_1=-2\veps_1=0.5$,
$\Gamma_1=0.05$, $\veps_2=0$, $\Gamma_2=0.02$, $\omega_0=0.1$,
$\lambda=0.2$, and nine values of $|h-h_c|$ where $h_c\simeq 0.284959$.
Properties at the transition (thick lines without symbols) are those
expected at a level crossing between the local-charge phase (filled symbols)
and the \ASCd\ phase. $N_s = 3\,000$ states were retained after each NRG
iteration.}
\end{figure}

\subsubsection{Strong bosonic coupling}

Figure \ref{fig:r=2:thermo_c} shows $T\chi_{s,\imp}$,
$\frac{1}{4}T\chi_{c,\imp}$, and $S_{\imp}$ vs $T$ at particle-hole symmetry
($\veps_1=-\half U_1$) for a strong bosonic coupling $\lambda=0.2$ and various
local magnetic fields $h$ straddling the critical value $h_c\simeq 0.284959$.
Here, in contrast to Fig.~\ref{fig:r=2:thermo_s}, $T\chi_{s,\imp}$ falls to
zero in both the local-charge phase and the \ASCd\ phase, signaling the
suppression of spin fluctuations at the impurity site. However, the flows of
$T\chi_{c,\imp}$ with decreasing temperature clearly reveal the existence of a
QPT separating the \ASCd\ and local-charge phases. Exactly at the critical
value $h=h_c$ (lines without symbols in Fig.\ \ref{fig:r=2:thermo_c}),
$\frac{1}{4}T\chi_{c,\imp}$ is pinned at low temperatures at the value $1/6$
expected at the point of degeneracy between the empty, spin-down, and doubly
occupied impurity configurations.
For $h$ deviating slightly from $h_c$, $T\chi_{c,\imp}$ traces the critical
behavior at high temperatures but eventually crosses below a scale $T^*_2$ to a
limiting value of either $1$ in the local-charge phase or $0$ in the \ASCd\
phase.

The temperature dependence of the spin (charge) susceptibility in
Fig.\ \ref{fig:r=2:thermo_c} mirrors that of the charge (spin) susceptibility
in Fig.~\ref{fig:r=2:thermo_s}. By contrast, the behavior of $S_{\imp}(T)$ is
equivalent in the two cases. These properties suggest that, as found for the
interacting quantum critical points for band exponents $0<r<1$ (Secs.\
\ref{sec:small-r_symm} and \ref{sec:small-r}), the quantum phase transitions
into/out of the LC phase at $\dd=0$ take place at points of enhanced symmetry
where the system acquires an SU(2) isospin invariance to match the global SU(2)
spin invariance of the Anderson-Holstein Hamiltonian in zero magnetic field.

Both at weak and strong bosonic couplings, the fact that the impurity properties
in the quantum-critical regime are those of the valence-fluctuation fixed point
(or its analog under interchange of spin and isospin) is entirely consistent
with the picture of each QPT as arising from a renormalized level crossing.
Moreover, crossover scales $T^*_1$ and $T^*_2$ extracted from the thermodynamic
properties are identical up to a constant multiplicative factor to those
identified from the NRG spectra (see Sec.\ \ref{subsec:r=2_symm:Tstar}).

\subsection{Linear conductance}
\label{subsec:cond}

It is generally impractical to measure the impurity thermodynamic properties
of a quantum-dot device. Rather, the primary experimental probe of lateral
quantum dots is electrical transport. The linear conductance of the
boson-coupled double-quantum-dot system modeled by Eq.\ \eqref{H_DD} can be
calculated from the Landauer formula
\begin{equation}
\label{cond}
g(T)=\frac{e^2}{h}\sum_{\s} \int^{\infty}_{-\infty} \! d\omega \:
  (-\partial f/\partial\omega)[-\text{Im}\mathcal{T}_{\s}(\omega)],
\end{equation}
where $f(\omega,T)=[\exp(\omega/T)+1]^{-1}$ is the Fermi-Dirac distribution
function and $\mathcal{T}_{\s}(\omega)=\pi\rho_0\sum_{i,j} V_i \,
G_{ij}^{\s}(\omega) \, V_j$ with
$G_{ij}^{\s}(\omega) = -i \int_0^{\infty} \! dt e^{i\omega t}
\langle \{ d_{i\s}^{\pdag}(t), d_{j\s}^{\dag}(0)\}\rangle$.
For $U_2=0$, one can re-express\cite{Dias:08}
\begin{align}
\label{Im T}
\! -\text{Im}\,\mathcal{T}_{\s}(\omega)
&= \bigl[ 1 - 2\pi\Gamma_2\rho_2(\omega) \bigr] \pi\Gam(\omega)
   A_{11}^{\s}(\omega) + \pi\Gamma_2\rho_2(\omega) \nonumber \\
&\quad + 2\pi(\omega-\veps_2)\Gam(\omega)\rho_2(\omega)\,
   \text{Re}\,G_{11}^{\s}(\omega),
\end{align}
where $\Gam(\omega)$ is as defined in Eq.\ \eqref{Gamma_dot},
$\rho_2(\veps)=(\Gamma_2/\pi)[(\veps-\veps_2)^2+\Gamma_2^2]^{-1}$ is a
Lorentzian of width $\Gamma_2$ centered on energy $\veps_2$, and
$A_{11}^{\s}(\omega)=-\pi^{-1}\text{Im}\,G_{11}^{\s}(\omega)$. All
quantities entering Eq.\ \eqref{cond} are known exactly with the sole exception
of $G_{11}^{\s}(\omega)$, the full dot-1 spin-$\s$ local Green's
function $G_{11}^{\s}(\omega)$ taking into account both electron-electron
($U_1$) and electron-boson ($\lambda$) coupling.

We have used standard NRG methods\cite{Bulla:08} to obtain the dot-1 spectral
function $A_{11}^{\s}(\omega)$ from  the effective one-impurity pseudogap
Anderson-Holstein model described by Eq.\ \eqref{H_DD:PAH}. After
$\text{Re}\,G_{11}^{\s}(\omega)$ has been obtained via the Kramers-Kronig
relations, Eqs.\ \eqref{cond} and \eqref{Im T} yield the linear conductance.
We present results only for zero magnetic field
[where $A_{11}^{\up}(\omega)=A_{11}^{\dn}(\omega)$] and/or for
strict particle-hole symmetry [where $A_{11}^{\up}(\omega)=
A_{11}^{\dn}(-\omega)$], special cases in which the up and down spin
channels contribute equally to the conductance. Temperatures are expressed as
multiples of $T_{K0}=7\times10^{-4}$, the Kondo temperature for the conventional
(i.e., metallic or $r=0$) one-impurity Anderson model with $U=-2\Ed=0.5$ and
$\Gamma=0.05$, which serves as a characteristic scale for the many-body physics
of the problem.

\begin{figure}[t]
\centering
\includegraphics[angle=270,width=0.95\columnwidth]{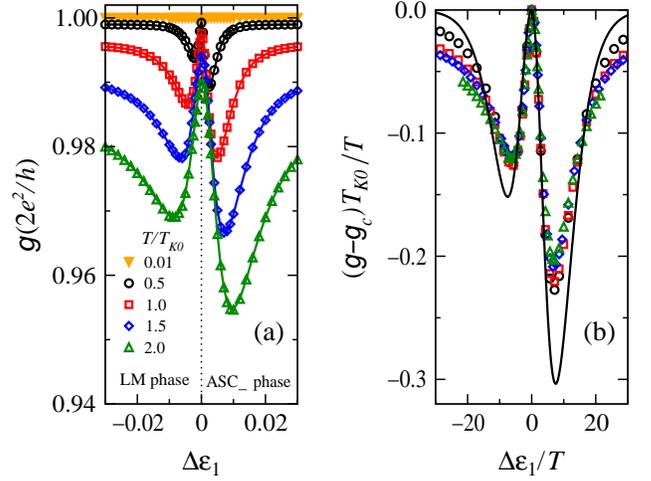}
\caption{\label{fig:cond_s}
(Color online) Linear conductance of a $U_2 = 0$ double-quantum-dot device near
a spin-sector QPT:
(a) Linear conductance $g$ vs $\Delta\veps_1=\veps_1-\veps_{1,c}^+$ for
$U_1=0.5$, $\Gamma_1=0.05$, $\veps_2=0$, $\Gamma_2=0.02$, $h=0$, $\omega_0=0.1$,
$\lambda=0.1$, and different temperatures $T$ specified in the legend as
multiples of $T_{K0}=7\times10^{-4}$. The retention of $N_s = 1\,000$ states
after each NRG iteration accounts for the small discrepancy between
$\veps_{1,c}^+\simeq-0.1249871$ and its value in the case $N_s = 3\,000$ shown in
Fig.\ \ref{fig:r=2:thermo_s}.
(b) The same data scaled as $(g-g_c)T_{K0}/T$ vs $\Delta\veps_1/T$, where
$g_c(T)$ is the conductance at $\veps_1=\veps_{1,c}^+$. The solid line was
obtained from Eq.\ \protect\eqref{cond1} by approximating $A_q(x)=a_q$ and
using values of $a_q$ and $\omega_q/(\veps_1-\veps_1^+)$ fitted from
$A_{11}^{\s}(\omega)$.}
\end{figure}

\subsubsection{Weak bosonic coupling}

Figure \ref{fig:cond_s}(a) plots $g$ versus $\Delta\veps_1=\veps_1-\veps_{1,c}^+$
for a weak bosonic coupling $\lambda=0.1$ and five temperatures $T$ listed in
the legend. At $T=0$, the linear conductance $g$ is structureless and takes
its maximum possible value $2e^2/h$, signaling perfect electron transaction
through the system. However, at $T>0$, $g$ versus $\Delta\veps_1$ develops clear
minima on either side of a maximum located precisely on the boundary
$\Delta\veps_1=0$ between the local-moment and \ASCm\ phases. The
peak-and-valley structure becomes more prominent upon increasing temperature
up to several times $T_{K0}$, making it  amenable to experimental observation.
Similar features have been reported [see Fig.\ 2(b) of Ref.\
\onlinecite{Dias:08}] for a double-quantum-dot system without bosonic coupling.

The essential features of the results shown in Fig.\ \ref{fig:cond_s}(a) can be
understood from the fact that---just as in the case of zero bosonic
coupling\cite{Dias:08}---near the QPT, the low-energy part of the dot-1
spectral function is dominated at low temperatures by a quasiparticle peak at
frequency $\omega_q \propto \veps_1 - \veps_1^+$.
Upon raising the temperature, this peak rapidly disappears once
$T\gtrsim\omega_q$. We approximate this behavior by
\begin{equation}
\label{A_11:qp}
A_{11}^{\s}(\omega)
  \simeq A_q(\omega_q/T) \; \delta(\omega-\omega_q)
  \quad \text{for } |\omega|\ll\Gamma_2 ,
\end{equation}
where $A_q(x)$ is an unknown scaling function that satisfies $A_q(x)\to 0$ for
$x\ll 1$ and $A_q(x)\to a_q > 0$ for $x\gg 1$.
Hilbert transformation of Eq.\ \eqref{A_11:qp} leads to
\begin{equation}
\label{Re G_11:qp}
\text{Re}\,G_{11}^{\s}(\omega,T)
  \simeq R_0(\omega_q,T) + R_1(\omega_q,T) \, \frac{\omega}{\Gamma_2} +
     \frac{A_q(\omega_q/T)}{\omega_q - \omega} ,
\end{equation}
where $R_0$ and $R_1$ are determined by the form of $A_{11}^{\s}(\omega)$
at $|\omega|\gtrsim\Gamma_2$, and may vary with $\veps_1$ and hence $\omega_q$.
Inserting Eqs.\ \eqref{A_11:qp} and \eqref{Re G_11:qp} into Eqs.\ \eqref{cond}
and \eqref{Im T}\ yields, for $\veps_2 = 0$ and $T \ll \Gamma_2$,
\begin{align}
\label{cond1}
g
&= \frac{2e^2}{h} \biggl\{ 1 -
   \frac{\pi^2}{3} \, \biggl( \frac{T}{\Gamma_2} \biggr)^2 \notag \\
& \quad - A_q(\omega_q/T) \, \frac{\Gamma_1 T}{\Gamma_2^2}
   \biggl[\frac{\pi (\omega_q/T)^2 \, e^{\omega_q/T}}
   {(e^{\omega_q/T} + 1)^2} \\
& \quad + \frac{4T}{\Gamma_2} \int_0^{\infty} \!\! dx \;
   \frac{x^4}{x^2 - (\omega_q/T)^2} \, \frac{e^x}{(e^x + 1)^2} \biggr]
   + O \biggl( \frac{T}{\Gamma_2} \biggr)^4 \biggr\} . \notag
\end{align}
The first line in Eq.\ \eqref{cond1}, which describes resonant tunneling through
dot 2, dominates the conductance both for $T\ll\omega_q$ and for $T\gg\omega_q$.
However, for $T\simeq\omega_q$, the conductance is dominated by the first term
in the square brackets, which arises from the $A_{11}^{\s}$ term in Eq.\
\eqref{Im T}. To good approximation, the conductance near the QPT
(where $\omega_q=0$) can be expressed as
\begin{equation}
\label{cond2}
g(\veps_1,T) = g(\veps_{1,c}^+,T)
  + T \, g_1\!\left(\frac{\veps_1 - \veps_{1,c}^+}{T} \right)
\end{equation}
with $g_1(0)=0$. Figure \ref{fig:cond_s}(b) shows that this form is
obeyed well by the NRG results. The precise scaling function $g_1$ cannot be
determined without knowledge of $A_q(x)$, but the zeroth-order approximation
$A_q(x)=a_q$ produces a reasonably good description of the numerical data.
This scaling collapse of the finite-temperature conductance feature provides a
clear signature of the underlying $T=0$ phase transition that may be sought in
experiments.

\subsubsection{Strong bosonic coupling}
As one would expect given the equivalence under spin-charge interchange of the
thermodynamic properties at the LM-\ASCpm\ and LC-\ASCud\ phase boundaries, the
variation of the conductance with $h$ around the critical field $h_c$ for
$\lambda>\lambda_0$ is very similar to the variation of $g$ with $\veps_1$
near $\veps_{1c}^{\pm}$. The system exhibits perfect electron transmission
($g=2e^2/h$) at $T=0$ and with increasing temperature develops an increasingly
prominent peak-and-valley signature of the
QPT. This signature can be understood as arising from the
existence of quasiparticle peaks $A_{11}^{\up}(\omega) \simeq
A_q(\omega_q/T) \, \delta(\omega-\omega_q)$ and $A_{11}^{\dn}(\omega)
\simeq \, A_q(\omega_q/T) \, \delta(\omega+\omega_q)$ at a frequency
$\omega_q \propto h - h_c$. Analysis similar to that applied in the case of
weak bosonic couplings leads to the prediction
\begin{equation}
\label{cond3}
g(h,T) = g(h_c,T) + T \, \tilde{g}_1\!\left(\frac{h-h_c}{T}\right) ,
\end{equation}
a scaling that is indeed displayed by the numerical data. (We do not show
these data explicitly due to their similarity with Fig.\ \ref{fig:cond_s}.)

\section{Summary}
\label{sec:summary}

We have conducted a study of the pseudogap Anderson-Holstein model describing a
magnetic impurity level that hybridizes with a pseudogapped fermionic host with
a density of states vanishing as $|\veps|^r$ at the Fermi energy ($\veps=0$),
and that is also coupled, via its charge, to a local-boson mode. The reduction
of the density of low-energy band excitations leads to quantum phase
transitions (QPTs) that can be classified into different types depending on the
strength of the impurity-boson coupling and on the presence or absence
of particle-hole and time-reversal symmetry. The main results are as follows:

(1) Under conditions of strict particle-hole and time-reversal symmetry,
the pseudogap Anderson-Holstein model with exponent $0<r<\half$ features two
types of continuous QPT. For a weak (strong) impurity-boson coupling that
results in a positive (negative) effective Coulomb interaction between electrons
in the impurity level, increasing the impurity-band hybridization from zero
drives the system through a continuous QPT between a local-moment
(local-charge) phase, in which a two-fold degree of freedom survives to $T=0$,
and a symmetric strong-coupling phase in which the impurity degree of freedom
is quenched by the conduction band. Critical exponents characterizing the
response to a local symmetry-breaking field suggest that these QPTs belong
to the same universality class as the QPT of the particle-hole-symmetric
pseudogap Anderson model.

(2) For $r\ge\half$, the symmetric strong-coupling fixed point is unstable
(just as in the pseudogap Anderson model without bosons) and for weak (strong)
impurity-boson couplings, a system exhibiting particle-hole and time-reversal
symmetry always lies in the local-moment (local-charge) phase.

(3) For weak impurity-boson couplings and away from particle-hole symmetry, the
symmetric strong-coupling phase is replaced by two asymmetric strong-coupling
phases, one corresponding to an empty impurity level and the other to double
occupation of the impurity site. These phases are separated from the
local-moment phase by QPTs in the same universality class as those of the
particle-hole-asymmetric pseudogap Anderson model. These QPTs are continuous
and interacting for $0<r<1$, but first order for $r\ge 1$.

(4) For strong impurity-boson couplings and in the presence of a magnetic field,
the local-charge phase is separated by QPTs (again in the asymmetric
pseudogap-Anderson universality class, and continuous for $r<1$ but first-order
for $r\ge 1$) from two asymmetric strong-coupling phases corresponding to single
occupation of the impurity level with either a spin-up or a spin-down electron.

(5) For $r=2$, the pseudogap Anderson-Holstein model provides a description of
two quantum dots connected in parallel to current leads, where one dot is tuned
to lie in a Coulomb blockade valley and is coupled via its charge to a
local-boson mode, while the other dot is tuned to be effectively noninteracting
and in resonance with the leads. The setup exhibits voltage- or
magnetic-field-tuned QPTs of the level-crossing type. These QPTs produce
peak-and-valley features in the linear conductance that become more prominent
upon increase of the temperature. Moreover, in the vicinity of the transitions,
the conductance data collapse to a single function of the ratio of a symmetry
breaking field to the absolute temperature.

\begin{acknowledgments}
We thank Luis G.\ G.\ V.\ Dias da Silva for valuable discussions.
Much of the computational work was performed at the University of Florida
High-Performance Computing Center. This work was supported in part by NSF
Grants No.\ DMR-0710540 and DMR-1107814.
\end{acknowledgments}

\appendix

\section{Derivation of Poor-Man's Scaling Equations}
\label{sec:scaling}

In this appendix, we outline the derivation of the poor-man's scaling equations
\eqref{U:scaling}--\eqref{Gamma:scaling} discussed in Sec.\ \ref{subsec:scaling}.
For this purpose, it proves convenient to work with the Anderson-Holstein
Hamiltonian in the form
\begin{equation}
\label{H_PAHM2}
\H' = \H'_{\imp} + \H_{\band} + \H_{\boson} + \H'_{\impband} + \H_{\impboson}\,,
\end{equation}
where $\H_{\band}$, $\H_{\boson}$, and $\H_{\impboson}$ are as defined in Eqs.\
\eqref{H_band}, \eqref{H_boson}, and \eqref{H_impboson}, respectively, but
$\H_{\imp}$ in Eq.\ \eqref{H_imp} is rewritten in more conventional fashion as
\begin{equation}
\label{H_imp2}
\H'_{\imp} = \Ed \, n_d + U \n_{d\up} \n_{d\dn}
\end{equation}
and $\H_{\impband}$ in Eq.\ \eqref{H_impband} is generalized to
\begin{align}
\label{H_impband2}
H'_{\impband}
&= \frac{1}{\sqrt{N_k}} \sum_{\bk,\s}
   \bigl\{ \bigl[ V_{0,\bk} (1-\n_{d,-\s}) \notag \\
&\qquad + V_{2,\bk} \, \n_{d,-\s} \bigr] d^{\dag}_{\s}
  c^{\pdag}_{\bk\s} + \text{H.c.} \bigr\},
\end{align}
where the hybridization functions
\begin{equation}
\label{Gamma_tau}
\Gam_{\tau}(\veps)=\frac{\pi}{N_k}\sum_{\bk} |V_{\tau,\bk}|^2
  \delta(\veps-\Ek) = \Gamma_{\tau}|\veps/D|^r \: \Theta(D-|\veps|)
\end{equation}
for $\tau = 0$, $2$ have the same power-law dependence as $\Gam(\veps)$
defined in Eq.\ \eqref{Gamma:power}. At the bare Hamiltonian level, one
expects the hybridization matrix element $V_{0,\bk}$ between the empty and singly
occupied impurity configurations to be identical to that $V_{2,\bk}$ between the
singly occupied and doubly occupied impurity configurations. However, this
degeneracy can be broken under the scaling procedure.

A canonical transformation $\H'\to \bH' = e^S \H' e^{-S}$ with $S$ as
defined in Eq.\ \eqref{Lang-Firsov} yields
\begin{equation}
\label{bar H_PAHM2}
\bH' = \bH'_{\imp} + \H_{\band} + \H_{\boson} + \bH'_{\impband} \, ,
\end{equation}
where $\bH'_{\imp}$ contains shifted parameters $\Ubar$
[Eq.\ \eqref{Ubar:def}] and $\Edbar$ [Eq.\ \eqref{Edbar:def}], and
\begin{align}
\label{bar H_impband2}
\bH'_{\impband}
&= \frac{1}{\sqrt{N_k}} \sum_{\bk,\s}
   \bigl\{ B^{\dag} \bigl[ V_{0,\bk} (1-\n_{d,-\s}) \notag \\
&\qquad + V_{2,\bk} \, \n_{d,-\s} \bigr] d^{\dag}_{\s}
  c^{\pdag}_{\bk\s} + \text{H.c.} \bigr\},
\end{align}
with $B$ as defined in Eq.\ \eqref{B:def}.

We analyze the problem using a basis of many-body states composed as direct
products of (i) fermionic states formed by the action of creation and
annihilation operators on $|FS\rangle$, the half-filled Fermi sea having $N_k$
electrons of energy $\Ek<0$, and (ii) occupation number eigenstates $|n)$ of
the transformed boson mode defined in Eq.\ \eqref{bar b:def}. Since real
occupation of states $|n)$ with $n>0$ is negligible in the anti-adibatic regime,
we focus on the states $|0,0\rangle=|FS\rangle\otimes|0)$,
$|\s,0\rangle=d_{\s}^{\dag}|0,0\rangle$, and
$|2,0\rangle=\sigma \, d_{\s}^{\dag}|\!-\!\s,0\rangle$.
Neglecting for the moment the effect of the hybridization [$\bH'_{\impband}$
in Eq. \eqref{bar H_impband2}], the energies of these states are denoted $E_0$,
$E_1=E_0+\Edbar$, and $E_2=E_1+\Edbar+\Ubar=2E_1-E_0+\Ubar$.

We now consider the effect of an infinitesimal reduction in the half-bandwidth
from $D$ to $\tD=D+dD$, where $dD < 0$. The goal is to write a new Hamiltonian
$\tilde{H}'$ similar in form to $\bH'$ but retaining only conduction-band
degrees of freedom having energies $|\Ek|<\tD$ and having parameters
$\tEd$, $\tU$, and $\tG_{\tau}$ adjusted to account perturbatively for the
band-edge states that have been eliminated.

Let $K^+$ be the set of wave vectors $\bk$ describing particle-like states
having energies $\tD<\Ek<D$, and $K^-$ be the set of wave vectors describing
hole-like state with energies $-D<\Ek<-\tD$. Tunneling of an electron from a
$K^-$ state into the empty impurity level, accompanied by the creation of
$n=0$, $1$, $\ldots$ local bosons, transforms the state $|0,0\rangle$ to
\begin{align}
\label{tilde0}
|\widetilde{0,0}\rangle
&= |0,0\rangle - \frac{e^{-\lambda^2/2\omega_0^2}}{\sqrt{N_k}}
  \sum_{\bk\in K^-,\,\s} V_{0,\bk} \\
&\times \sum_{n=0}^{\infty} \frac{1}{\sqrt{n!}} \, \frac{(\lambda/\omega_0)^n}
  {|\Ek|+E_1-E_0+n\omega_0} \; c_{\bk\s} |\s,n\rangle +O(V^2)
  \notag
\end{align}
with energy
\begin{align}
\label{tildeE_0:1}
\tE_0
&= E_0 - \frac{e^{-\lambda^2/\omega_0^2}}{N_k} \sum_{\bk\in K^-,\,\s}
   |V_{0,\bk}|^2 \notag \\
&\qquad \times
  \sum_{n=0}^{\infty} \frac{1}{n!} \, \frac{(\lambda/\omega_0)^{2n}}
   {|\Ek|+E_1-E_0+n\omega_0} + O(V^3) .
\end{align}
Here, $O(V^m)$ schematically represents all processes involving at
least $m$ factors $V_{\tau_1,\bk_1} \cdots V_{\tau_n,\bk_n}$.
The derivation of Eqs.\ \eqref{tilde0} and \eqref{tildeE_0:1} makes use of
\begin{align}
(n|e^{\pm\alpha(b^{\dag}-b)}|0)
&= (n|e^{\pm\alpha(\bar{b}^{\dag}-\bar{b})}|0) \notag \\
&= e^{-\alpha^2/2} \,
  (n|e^{\pm\alpha\bar{b}^{\dag}}e^{\mp\alpha \bar{b}}|0) \notag \\
&= \frac{e^{-\alpha^2/2}}{\sqrt{n!}} \, (\pm \alpha)^n .
\end{align}
Since $N_k^{-1} \sum_{\bk\in K^{\pm}} V_{\tau,\bk}^2 \simeq \pi^{-1}
\Gam_{\tau}(\pm D)\,\delta(\Ek\mp D)$, one can re-express the perturbed energy
\begin{equation}
\label{tildeE_0}
\tE_0 \simeq E_0 - |dD| \, \frac{2\Gam_0(-D)}{\pi\,\E(D\!+\!\Edbar)} + O(V^3),
\end{equation}
where $\E(E)$ is the energy function defined in Eq.\ \eqref{E:def}.

Similarly, tunneling of an electron from the doubly occupied impurity level
into a $K^+$ state transforms $|2,0\rangle$ to
\begin{align}
\label{tilde2}
|\widetilde{2,0}\rangle
&= |2,0\rangle - \frac{e^{-\lambda^2/2\omega_0^2}}{\sqrt{N_k}}
   \sum_{\bk\in K^+,\,\s} V_{2,\bk} \\
&\times \sum_{n=0}^{\infty} \frac{1}{\sqrt{n!}} \, \frac{(\lambda/\omega_0)^n}
   {\Ek-\Ubar-\Edbar+n\omega_0} \; c^{\dag}_{\bk\s} |\!-\!\s,n\rangle
   + O(V^2) \notag
\end{align}
with energy
\begin{align}
\label{tildeE_2}
\tE_2
&= E_2 - \frac{e^{-\lambda^2/\omega_0^2}}{N_k} \sum_{\bk\in K^+,\,\s}
  |V_{2,\bk}|^2 \notag \\
&\quad \times
  \sum_{n=0}^{\infty} \frac{1}{n!} \, \frac{(\lambda/\omega_0)^{2n}}
  {\Ek-\Ubar-\Edbar+n\omega_0} +O(V^3) \notag \\
&\simeq E_2 - |dD| \, \frac{2\Gam_2(D)}{\pi\,\E(D\!-\!\Ubar\!-\!\Edbar)} \
  + O(V^3).
\end{align}
Finally, tunneling of an electron into the singly occupied impurity from a
$K^-$ state or from the singly occupied level into a $K^+$ state transforms
$|\s,0\rangle$ to
\begin{align}
\label{tilde1}
|\widetilde{\s,0}\rangle
&= |\s,0\rangle - \frac{e^{-\lambda^2/2\omega_0^2}}{\sqrt{N_k}}
   \left[ \: \sum_{\bk\in K^-} V_{2,\bk} \right. \notag \\
&\times
   \sum_{n=0}^{\infty} \frac{1}{\sqrt{n!}} \, \frac{(\lambda/\omega_0)^n}
   {|\Ek|+E_2-E_1+n\omega_0} \; c_{\bk,-\s} |2,n\rangle \notag \\
&\left. - \!\! \sum_{\bk\in K^+} \!\! V_{0,\bk}
   \sum_{n=0}^{\infty} \frac{1}{\sqrt{n!}} \, \frac{(\lambda/\omega_0)^n}
   {\Ek+E_0-E_1+n\omega_0} \; c^{\dag}_{\bk\s} |0,n\rangle \right]
   \notag \\
&\qquad +O(V^3)
\end{align}
with energy
\begin{align}
\label{tildeE_1}
\tE_1
&= E_1 - \frac{e^{-\lambda^2/\omega_0^2}}{N_k} \left[ \: \sum_{\bk\in K^-}
   |V_{2,\bk}|^2 \right. \notag \\
&\times \sum_{n=0}^{\infty} \frac{1}{n!} \, \frac{(\lambda/\omega_0)^{2n}}
   {|\Ek|+E_2-E_1+n\omega_0} \notag \\
&\left. - \sum_{\bk\in K^+} |V_{0,\bk}|^2
   \sum_{n=0}^{\infty} \frac{1}{n!} \, \frac{(\lambda/\omega_0)^{2n}}
   {\Ek+E_0-E_1+n\omega_0} \right] +O(V^3) \notag \\
&\simeq E_1 - |dD|\left[ \frac{\Gam_2(-D)}{\pi\,\E(\!D+\!\Ubar\!+\!\Edbar)}
   + \frac{\Gam_0(D)}{\pi\,\E(D\!-\!\Edbar)} \right] \notag \\
&\qquad + O(V^3).
\end{align}
The $O(V^2)$ terms in each of the above states $|\widetilde{\phi,0}\rangle$
include terms to enforce normalization, i.e.,
$\langle\widetilde{\phi,0}|\widetilde{\phi,0}\rangle =
\langle\phi,0|\phi,0\rangle = 1$.

The modified energies can be used to define effective Hamiltonian parameters
$\tEd=\tE_1-\tE_0$ and $\tU=\tE_2+\tE_0-2\tE_1$. At the same time, for each
$\bk$ in the retained portion of the band (i.e., satisfying $|\Ek|<\tD$),
$V_{0,\bk}$ must be replaced by
\begin{equation}
\tilde{V}_{0,\bk} =
  \begin{cases}
    \sqrt{N_k} \: \langle \widetilde{0,0}|B^{\dag} c_{\bk\s} \H'
      |\widetilde{\s,0}\rangle
    & \text{for } \Ek > 0 \\[1ex]
    - \sqrt{N_k} \: \langle \widetilde{\s,0}| B c_{\bk\s}^{\dag} \H'
      |\widetilde{0,0}\rangle
    & \text{for } \Ek < 0 ,
  \end{cases}
\end{equation}
and $V_{2,\bk}$ must be replaced by
\begin{equation}
\tilde{V}_{2,\bk} =
  \begin{cases}
  -\sigma \sqrt{N_k} \: \langle\widetilde{\s,0}|B^{\dag} c_{\bk,-\s} \H'
    |\widetilde{2,0}\rangle
  & \text{for } \Ek > 0 \\[1ex]
  \sigma \sqrt{N_k} \: \langle \widetilde{2,0}| B c_{\bk,-\s}^{\dag} \H'
    |\widetilde{\s,0}\rangle
  & \text{for } \Ek < 0 .
  \end{cases}
\end{equation}
It is straightforward to show that
\begin{equation}
\label{tildeV_tau,k}
\tilde{V}_{\tau,\bk} = V_{\tau,\bk} + O(V^3).
\end{equation}
We shall not attempt to evaluate the leading corrections, which turn out
to be negligible in pseudogap ($r>0$) cases.

The infinitesimal band-edge reduction described in the previous paragraphs
can be carried out repeatedly to reduce the half-bandwidth by a finite
amount from $D$ to $\tD<D$. Equations \eqref{tildeE_0} and \eqref{tildeE_1}
indicate that during this process, the impurity level energy evolves
according to the scaling equation
\begin{equation}
\frac{d\tEd}{d\tD} = \frac{1}{\pi} \biggl[ \frac{\tG_{0,+}}{\E(\tD\!-\!\tEd)}
   - \frac{2\tG_{0,-}}{\E(\tD\!+\!\tEd)}
   + \frac{\tG_{2,-}}{\E(\tD\!+\!\tU\!+\!\tEd)} \, \biggr] + O(V^3),
\end{equation}
where $\tG_{\tau,\pm}$ is the value of the rescaled hybridization function at
the reduced band edges $\veps=\pm\tD$. Taking into account Eq.\
\eqref{tildeE_2} as well, one sees that the on-site repulsion follows
\begin{align}
\frac{d\tU}{d\tD}
&= \frac{2}{\pi} \biggl[ \frac{\tG_{0,+}}{\E(\tD\!+\!\tEd)}
   - \frac{\tG_{0,+}}{\E(\tD\!-\!\tEd)}
   + \frac{\tG_{2,+}}{\E(\tD\!-\!\tU\!-\!\tEd)} \notag \\
&\qquad - \frac{\tG_{2,-}}{\E(\tD\!+\!\tU\!+\!\tEd)} \, \biggr] + O(V^3).
\end{align}
The band-edge hybridization functions $\tG_{\tau}$ rescale both due to the
replacement of $D$ by $\tD$ in Eq.\ \eqref{Gamma:power} and due to the
perturbative corrections to $V_{\tau,\bk}$ in Eq.\ \eqref{tildeV_tau,k}, leading
to the scaling equation
\begin{equation}
\label{Gamma_tau:scaling}
\frac{d\tG_{\tau,\pm}}{d\tD} = r\,\frac{\tG_{\tau,\pm}}{\tD} + O(V^4) .
\end{equation}

The bare hybridization functions specified in Eq.\ \eqref{Gamma_tau} are such that
$\tG_{\tau,\pm}(D)=\Gamma$. For $r>0$, moreover, Eq.\ \eqref{Gamma_tau:scaling}
shows that the band-edge hybridization functions are irrelevant (in the RG sense), and so
too must be any differences among the renormalized values of the four hybridization
widths. It is therefore an excellent approximation to set $\tG_{\tau,\pm}=\tG$ from
the outset, leading to the simplified scaling equations given in
Eqs.\ \eqref{U:scaling}--\eqref{Gamma:scaling}.


\begin{thebibliography}{66}
\bibitem{H_imp:alt}
Equation \protect\eqref{H_imp} is equivalent, apart from an additive constant
$-\Ed$, to the more conventional form Eq.\ \protect\eqref{H_imp2}. The form of
Eq.\ \protect\eqref{H_imp} emphasizes the charge-conjugation symmetry of the
full Hamiltonian [Eq.\ \protect\eqref{H_PAHM}] in the special case $\dd=0$.
\bibitem{impurity-props}
The impurity contribution to a physical property is defined in the standard
manner (see, e.g., Ref.\ \protect\onlinecite{Krishna-murthy:80}) as
the difference between the total value of the property for the system
including the impurity and the total value of the same property for the
host alone. In the case of the Anderson-Holstein model (with or without a
pseudogap), we treat the local boson mode as part of the host. As a result,
the impurity entropy takes a finite value $S_{\imp}=\ln 4$ in the
high-temperature limit.
\bibitem{SSC-mag-field}
Extension of the analysis in Sec.\ IV of Ref.\
\protect\onlinecite{Gonzalez-Buxton:98} indicates that in the vicinity of
the symmetric strong-coupling fixed point of the pseudogap Anderson model,
breaking time-reversal symmetry admits a relevant perturbing operator
$O_{H_1} = \half \Lambda^{N/2} (f_{1\up}^{\dag} f_{1\up}^{\pdag}
- f_{1\dn}^{\dag} f_{1\dn}^{\pdag}) \propto T^{-\min(r,1)}$
where the $f_{n\s}$ is one of the annihilation operators entering
Eq.\ \protect\eqref{tight-binding}. By contrast, the leading perturbation
introduced near the asymmetric strong-coupling fixed point is an irrelevant
operator $O_{H_2} = \half \Lambda^{N/2} (f_{2\up}^{\dag} f_{2\up}^{\pdag}
- f_{2\dn}^{\dag} f_{2\dn}^{\pdag}) \propto T^r$.
\bibitem{Fritz-note}
This conclusion contradicts Ref.\ \onlinecite{Fritz:04} but is supported by
numerical renormalization-group calculations (Ref.\ \onlinecite{Cheng:inprep}).
\bibitem{e_n}
For any situation that does not obey
$\tG(-\veps)\ne\tG(\veps)$ for all $\veps$, the right-hand side of
Eq.\ \protect\eqref{tight-binding} must be augmented by a term
$\sum_{n,\s} \Lambda^{-n/2} e_n f_{n\s}^{\dag} f_{n\s}^{\pdag}$,
where the on-site energies $e_n$ parametrize the particle-hole asymmetry
of the hybridization function.
\end{thebibliography}
\end{document}